\documentclass{aa}

\usepackage{graphicx}
\usepackage{caption}
\usepackage{txfonts}

\usepackage[colorlinks=true, allcolors=blue]{hyperref}
\usepackage{caption}

\newcommand{\kms}{km~s$^{-1}$}

\def \HH{\ifmmode{\rm H_2}\else{$\rm H_2$}\fi}
\def \ecqs  {\ifmmode{\,{\rm erg}\,{\rm cm}^{-2}\,{\rm s}^{-1}\,{\rm 
             sr}^{-1}}\else{$\,{\rm erg}\,{\rm cm}^{-2}\,{\rm s}^{-1}\,{\rm sr}^{-1}$}\fi}

\usepackage{overpic}
\usepackage{multicol, multirow}
\usepackage{longtable}

\begin{document}

\title{Turning JWST/MIRI backgrounds into a survey of \\
diffuse molecular hydrogen}

\author{
E. Nigou\inst{\ref{lpens}, \ref{obs}}  \and
B. Godard\inst{\ref{lpens},\ref{obs}}  \and
P. Guillard\inst{\ref{iap}}  \and
G. Pineau des Forêts\inst{\ref{obs}, \ref{ias}} \and
M.~A. Miville-Deschênes\inst{\ref{lpens},\ref{obs}} \and
P. Lesaffre\inst{\ref{lpens},\ref{obs}}
}
        
\institute{
Laboratoire de Physique de l’École Normale Supérieure, ENS, Université PSL, CNRS, Sorbonne Université, 75005 Paris, France \label{lpens} \and
LUX, Observatoire de Paris, Université PSL, Sorbonne Université, CNRS, 75014 Paris, France \label{obs} \and
Institut d'Astrophysique de Paris, Sorbonne Universit\'{e}, CNRS, 98\,bis bd Arago, 75014 Paris, France \label{iap} \and
Institut d’Astrophysique Spatiale, Université Paris-Saclay, CNRS, 91405 Orsay, France \label{ias}
}

\date{Received 26 January 2026 / Accepted 13 March 2026}

\abstract 
	{A statistically significant sampling of \HH\ rotational excitation in the diffuse interstellar medium (ISM) is essential to identifying its excitation mechanisms and assessing the importance of \HH\ in the cooling of the gas and the regulation of thermal pressure.}
    {To complement the statistics provided by ancillary telescopes, we conducted a search for pure rotational \HH\ emission lines in all publicly available background observations obtained with the Medium Resolution Spectrometer (MRS) aboard the JWST.}
    {The sample consists of 276 background observations acquired over the past three years. Departing from the standard pipeline, each uncalibrated MRS background file was reprocessed, enabling the analysis of \HH\ pure rotational emission.
    Lines of sight likely associated with star-forming complexes were excluded to focus on emission from the diffuse ISM. The results were compared with FUSE absorption data and were analyzed in relation to the column densities of H and \HH\ and to dust emission derived from HI4PI, \textit{Planck}, and WISE data.}
    {This analysis reveals widespread \HH\ emission throughout the Galaxy. We report the first detections of the pure rotational S(4), S(5), and S(7) lines in the diffuse ISM. The S(1) line is detected along 84 lines of sight, corresponding to a detection rate of 41\%. Its integrated intensity decreases steeply with Galactic latitude, spanning nearly two orders of magnitude, in remarkable agreement with absorption measurements. The $T_{34}$ and $T_{35}$ excitation temperatures vary between 200 and $\sim$1000~K, are correlated with each other, and are anticorrelated with the column density of \HH, as expected from ancillary data. All lines of sight in the sample have undergone the H--\HH\ transition, at $N_{\rm{H}} \gtrsim 10^{20} \ \rm{cm}^{-2}$, and are partly molecular, with $f_{\HH} \gtrsim 0.1$. Under these conditions, the cooling rate associated with the S(1) line, expressed per hydrogen atom, is found to be remarkably constant, with a characteristic value of $\sim 4\times10^{-27}$ erg s$^{-1}$ H$^{-1}$.}
	{This study demonstrates that the high sensitivity of the JWST enables measurements that both strengthen and complement those from absorption studies. Observations collected over just a fraction of JWST’s lifetime have already yielded detections along dozens of lines of sight, significantly expanding the statistical sample of \HH\ rotational excitation in the diffuse ISM.}


 \titlerunning{Turning JWST/MIRI backgrounds into a survey of diffuse molecular hydrogen}
 \maketitle

\section{Introduction}

\begin{figure*}[!t]
    \centering
    \includegraphics[width=1.0\linewidth,trim = 0.0cm 0.0cm 0.0cm 0.0cm,]{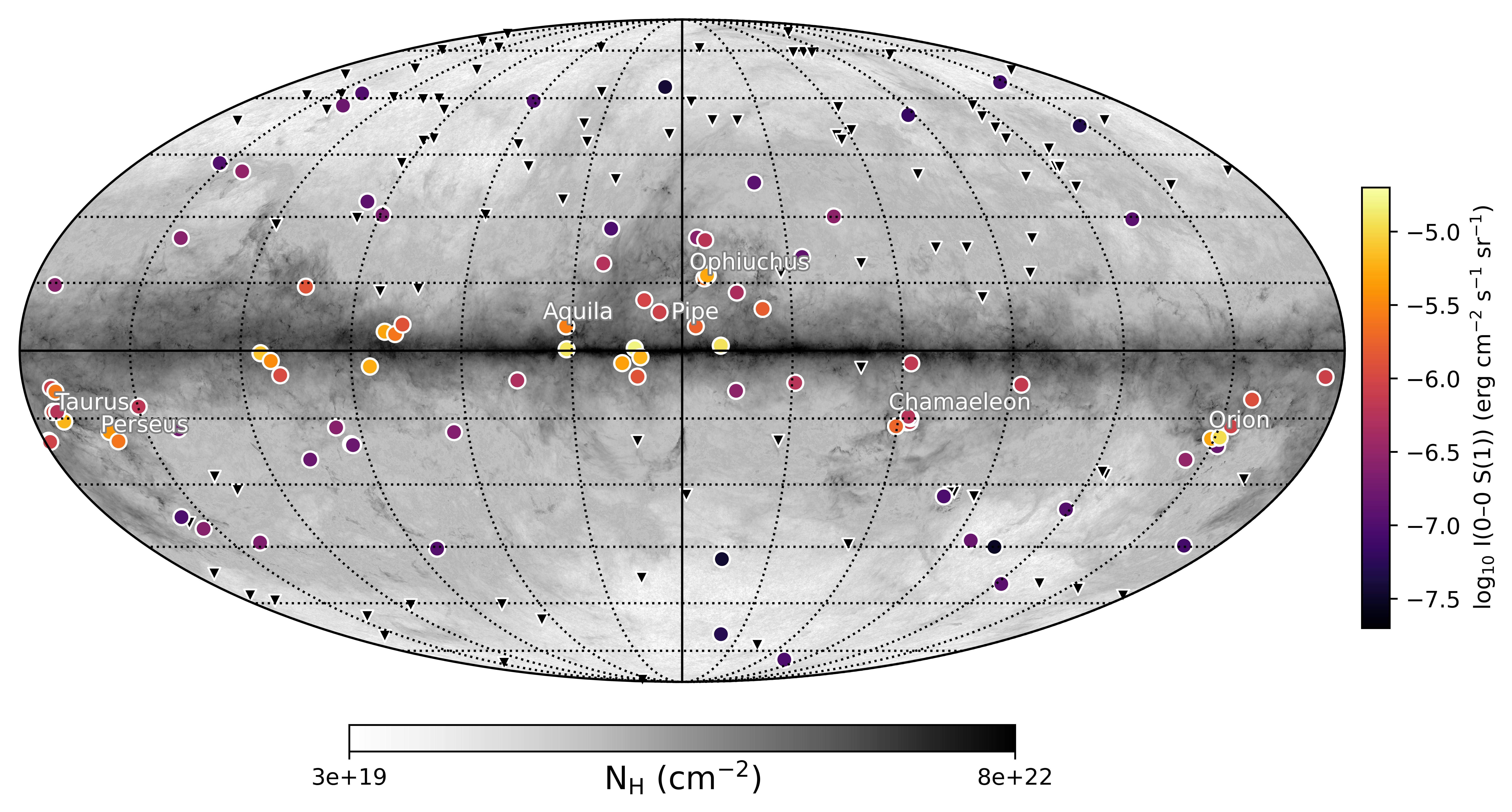}
    \caption{Aitoff projection of the MRS background lines of sight analyzed in this paper. Filled circles indicate detections of the 0--0 S(1) line of \HH, color-coded by integrated intensity, while triangles show non-detections. All lines of sight are overlaid on the total hydrogen column density map derived from the dust opacity at 353 GHz measured by \textit{Planck} \citep{Planck2014}.}
    \label{fig:aitoff}
\end{figure*}

Understanding the formation, survival, and excitation of molecular hydrogen is a long-standing objective in studies of the interstellar medium (ISM). First, \HH\ surface density correlates with the star formation rate \citep[e.g.,][]{Lin2019a} and  provides a valuable tracer for understanding the Kennicutt--Schmidt relation and its spatial variations in galaxies \citep[e.g.,][]{Leroy2013a, Pessa2022a}. Second, as the most abundant molecule in the Universe, \HH\ initiates interstellar chemistry and the formation of complex species in space. Last, while the cooling of the two stable phases of the diffuse ISM, the warm neutral medium (WNM) and the cold neutral medium (CNM), is presumably dominated by atomic species \citep{Wolfire2003}, \HH\ plays a subtle but essential role in cooling the gas. By preventing the CNM from evaporating into the WNM, \HH\ cooling suppresses thermal instability and regulates the thermal pressure of the diffuse gas and, hence, the relative distributions of the WNM and CNM \citep{Godard2023}.

The first detection of molecular hydrogen in the local diffuse ISM was obtained with UV rocket observations \citep{Carruthers1970} and was later confirmed and extensively characterized by the {\it Copernicus} satellite via resonant electronic absorption lines of \HH\ toward nearby OB stars \citep[e.g.,][]{Spitzer1975, Savage1977, Shull1982}. These early studies were later expanded upon using measurements by the Far Ultraviolet Spectroscopic Explorer (FUSE), which targeted more distant and fainter sources both in the Galactic disk \citep[e.g.,][]{Rachford2002, Shull2021a} and at higher Galactic latitudes toward extragalactic background sources \citep[e.g.,][]{Gillmon2006, Gillmon2006b, Wakker2006a}. The first detection of \HH\ pure rotational emission from diffuse Galactic regions came with the Infrared Space Observatory (ISO), which measured the S(0) to S(3) lines along a line of sight across the Galactic midplane \citep{Falgarone2005}. This discovery was later extended with {\it Spitzer} observations of the first pure rotational lines of \HH\ in emission toward two translucent clouds at high Galactic latitudes associated with infrared cirrus emission \citep{Ingalls2011a}.

On the one hand, observations performed in absorption have provided access to the so-called H--\HH\ transition and, more importantly, to the statistical distribution of the molecular fraction along random lines of sight. In the local diffuse ISM, this transition occurs at total hydrogen column densities $N_{\rm H} \sim 10^{20}-10^{21}$~cm$^{-2}$, with a mean transition column density that depends on the Galactic latitude \citep{Gillmon2006}. Across this range, the molecular fraction increases by several orders of magnitude and approaches a mean value of $f_{\rm \HH}\gtrsim 0.1$, with decreasing dispersion at higher column densities, reflecting line-of-sight averaging over long distances. Initially interpreted as a signature of H-to-\HH\ conversion within individual clouds, this transition is now understood to reflect the underlying multiphase structure of the gas \citep{Bellomi2020}. The distribution of \HH\ column densities traces the probability of encountering CNM structures along a given line of sight, which itself results from thermal instability and is ultimately regulated by Galactic-scale properties such as the mean density of the diffuse ISM, the intensity of the ambient UV radiation field, and the typical separation between OB associations.

On the other hand, observations of \HH\ in both emission and absorption have provided access to its excitation in its pure rotational levels. While the lowest two rotational states ($J=0,1$) appear to be thermally coupled with the gas at a kinetic temperature close to that of the CNM ($\sim 80$~K), higher rotational states display excitation temperatures of a few hundred Kelvin that anticorrelate with the column density of molecular hydrogen \citep{Spitzer1974, Savage1977, Wakker2006a, Shull2021a}. Two competing processes are invoked to account for the excitation of these rotational states: collisions in warm environments and far-UV pumping of the electronic transitions followed by fluorescent decay. Multiple studies have shown that the populations of several rotational levels cannot be explained by far-UV pumping alone \citep{Gry2002, Falgarone2005, Nehme2008, Ingalls2011a} and instead require collisional excitation in gas at temperatures of a few hundred Kelvin along the line of sight. This warm \HH\ component, which is connected to the presence of CH$^+$ in the diffuse ISM \citep[e.g.,][]{Frisch1980,Lambert1986}, may arise from the dissipation of mechanical energy \citep[e.g.,][]{Godard2009, Godard2014, Myers2015, Lesaffre2020, Moseley2021} or from the evaporation of CNM structures \citep{Valdivia2016,Godard2023,Godard2025a}. In this context, assembling a large observational sample is essential to enable statistical analyses of \HH\ excitation and to constrain models of multiphase, turbulent gas, as such statistics provide insight far beyond what can be inferred from individual lines of sight.

The present study aims to expand the statistical sampling of \HH\ rotational excitation in the diffuse ISM by searching for pure rotational line emission in all available background observations collected by the Medium Resolution Spectrometer (MRS) aboard the {\it James Webb} Space Telescope (JWST). This approach potentially enables the detection of \HH\ from the S(1) to S(8) transitions.

\section{JWST MIRI-MRS data}
\label{Sect2:Data}

\subsection{MRS backgrounds and data reduction}
\label{Sect:Datared}

The recommended observation strategy for JWST/MRS programs is to measure the background level using dedicated, co-temporal observations of a nearby region in the sky selected to be free of target emission. These background observations are primarily used to remove the sky and telescope contributions (including zodiacal light and thermal emission\footnote{See the latest \href{https://www.stsci.edu/files/live/sites/www/files/home/jwst/documentation/technical-documents/_documents/JWST-STScI-008978.pdf}{JWST technical report} for details.}), and to mitigate detector artifacts such as temporal variations in the hot and warm pixels and coherent striping produced by drifts in the effective detector dark current that are currently not captured by the standard calibration pipeline\footnote{While the pipeline corrects for dark current offsets in the dark count rate it cannot compensate for changes in the coherent pattern noise.}. Such detector systematics are best calibrated using dedicated backgrounds with the same readout ramp length as the science data.

Dithering offsets the target on the detector between exposures, improving spatial sampling and helping reject bad pixels and cosmic-ray residuals when the exposures are combined. A large fraction of MRS backgrounds used in this paper are taken with a two-point extended source dither pattern with detector readout mode and parameters (including exposure time set by the $N_{\rm groups}$ parameter) matched to that used for the science exposures. This is, in most cases, sufficient to mitigate the effects of bad pixels and cosmic rays \citep{Law2025}. In other cases, the data consist of either a single background position (i.e, no dithering) or a four-point extended dither pattern, with a total exposure depth comparable to that of the science observation, as typically used for very faint targets. The MRS covers a total wavelength range from $4.9 - 27.9\,\mu $m, separated into four integral field units (IFUs) referred to as channels, each divided into three bands. The channels cover slightly different fields of view (FoVs), from 3.2\arcsec$\times$3.7\arcsec (channel 1) up to 6.6\arcsec$\times$7.7\arcsec (channel 4), and have different spectral resolutions \citep[from $\sim$3700 to $\sim$1500,][]{Labiano2021}. 

Dedicated backgrounds (i.e., indicated as such in the Astronomer Proposal Tool) are generally not used directly for scientific purposes, and they are automatically subtracted from science observations in the pipeline. Therefore, for our purpose, we re-reduced the uncalibrated background MRS files up to level3 calibrated spectral cubes, the standard analysis-ready data product for IFU observations. The data reduction was done with the JWST Science Calibration Pipeline \citep[version 1.17.1,][]{Bushouse_1.17.1}, with the context 1350 for the Calibration References Data System following the standard procedures (see, e.g., \citealt{labiano_miri_2016} and \citealt{Alvarez2023} for detailed examples of MRS data reduction and calibration). 
Careful examination of the data showed that the stage 1 corrections \citep{Morrison2023} could be run with default parameters, and did not reveal significant residual impacting line emission. Naturally, we switched off the master background correction and sky matching steps in stage 3 of the pipeline, before producing the final fully reconstructed science cubes \citep{Law2023}. We did not realign the MRS astrometry thanks to the simultaneous imaging data. Finally, the science cubes were shifted to the usual orientation with north up and east to the left, increasing the total FoV in channel 1 to 6.1\arcsec$\times$5.6\arcsec and up to 11.5\arcsec$\times$11.5\arcsec in channel 4. 

\subsection{Lines of sight selection} 
\label{Sect:selection}

The data reduction described above was applied to all publicly available Mid-InfraRed Instrument-MRS (MIRI-MRS) observations retrieved from the \href{https://mast.stsci.edu/portal/Mashup/Clients/Mast/Portal.html}{Mikulski Archive for Space Telescopes (MAST)}. As of November 2025, this resulted in 297 individual lines of sight, each observed in at least one of the four spectral channels of the MRS instrument.  Among these, 21 lines of sight were excluded from the sample. Programs targeting bodies in the Solar System were rejected due to contamination of the background spectra by diffuse light scattered from the primary target \citep[e.g.,][]{Bockelee-Morvan2024}, while lines of sight toward the Galactic center were excluded because their mid-infrared emission is dominated by dense environments rather than by the diffuse ISM.

The spectral properties of the final sample, which comprises 276 lines of sight, and the procedures used to extract and identify \HH\ rotational emission are described in Appendix~\ref{append:data}. The corresponding measurements are summarized in Tables~\ref{tab:all-detect} and \ref{tab:statistics}, which report, respectively, the integrated intensities measured for all detected \HH\ rotational lines in each spectrum and the detection statistics for each transition. To support the interpretation of the data, we also compiled estimates of the atomic and total hydrogen column densities, as well as the dust temperature and its 12~$\mu$m emission, along each line of sight using all-sky surveys. These complementary data are discussed in Appendix~\ref{append:ancillary} and listed in Table~\ref{tab:ancillary}. Although these estimates are derived over solid angles at least 25 times larger than the MRS FoV, they provide valuable contextual information.

A subset of the final sample is displayed in Fig.~\ref{fig:aitoff}, which shows the distribution of lines of sight in Galactic coordinates, as well as the integrated intensity of the \HH\ 0--0 S(1) line, used throughout this paper as a reference for the study of warm and diffuse molecular hydrogen. Several background lines of sight exhibit rotational \HH\ emission significantly higher than the bulk of the sample. Upon further inspection, a fraction of these highly emissive lines of sight are associated with the Orion, Ophiuchus, Perseus, and Chamaeleon cloud complexes, all located near regions of active star formation. To exclude such environments and focus on the diffuse ISM, we devised selection criteria based on the total hydrogen column density and the dust temperature, as described in Appendix~\ref{append:flag}. Applying these criteria led to the flagging of 43 observations as non-diffuse, leaving a final sample of 233 lines of sight intersecting the diffuse ISM.

\section{Diffuse \HH\ emission and excitation}
\label{Sect:results}

We report widespread detections of \HH\ pure rotational line emission originating from Galactic material. After excluding the flagged lines of sight, the S(1) line is detected along 84 lines of sight, up to Galactic latitudes of $78^\circ$. Most detections correspond to molecular hydrogen column densities between $10^{19}$ and $10^{22}$~cm$^{-2}$, with line intensities and excitation properties consistent with absorption measurements in the solar neighborhood. This consistency indicates that the emission predominantly arises from diffuse, largely local interstellar gas\footnote{We note that two lines of sight have \HH\ column densities above $10^{22}$~cm$^{-2}$. Those lie at Galactic latitudes $|b|$ below $\sim 0.5^\circ$ and intersect the full column of Galactic disk material, similarly to the line of sight analyzed by \citet{Falgarone2005}.}. In this context, we also report the first detections of the pure rotational S(4), S(5), and S(7) lines in the local diffuse ISM.
As the detection rate generally decreases with increasing rotational quantum number (see Table~\ref{tab:statistics}), the S(6) and S(8) lines are not detected.
The spontaneous Einstein coefficient of the $J \rightarrow J-2$ pure rotational transition approximately scales as $J^6$ between $J=2$ and $J=9$ \citep{Shull1978}. This decrease in detection rate therefore reflects the sharp drop in the level populations as $J$ increases.

\subsection{Variations with galactic latitude}

\begin{figure*}
    \centering
    \includegraphics[width=17cm, trim = 0.0cm 0.0cm 0.0cm 0.0cm]{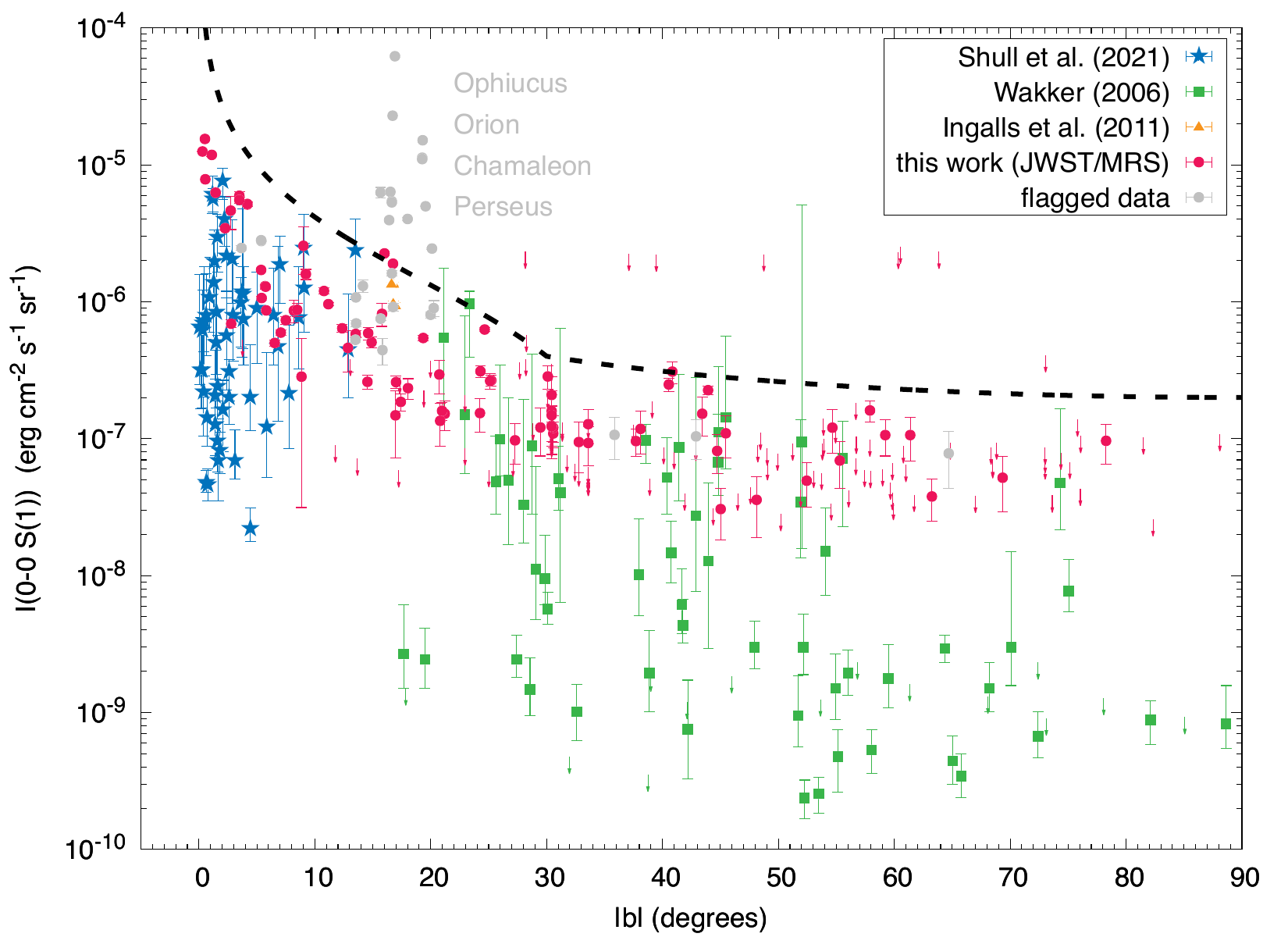}
    \caption{Integrated intensities of the 0--0 S(1) line of \HH\ as a function of the absolute Galactic latitude, $|b|$. Green squares and blue stars correspond to values inferred from absorption measurements toward extragalactic targets and nearby OB stars observed with FUSE \citep{Wakker2006a, Shull2021a}, using Eq.~\ref{Eq:cdint} and the column densities of the $J=3$ rotational level. Yellow triangles show measurements obtained with \textit{Spitzer} by \citet{Ingalls2011a}. Gray and red points represent intensities measured along flagged and un-flagged MRS background lines of sight, respectively, while red arrows indicate upper limits set by the sensitivity of the MRS over the wide range of exposure times in the sample. The dashed line shows a geometrically motivated analytical estimate of the maximum expected emission (Eq.~\ref{Eq:max}).}
    \label{fig:s1vsb}
\end{figure*}

The integrated intensities of the \HH\ 0--0 S(1) line are shown in Fig.~\ref{fig:s1vsb} as functions of the absolute Galactic latitude of the lines of sight, $|b|$. The measured intensities span more than two orders of magnitude and display a steep decline with increasing latitude, in excellent agreement with values inferred from absorption studies. For exposure times of a few minutes—which apply to most MRS background pointings—the sensitivity limit is $\sim 10^{-7}$~\ecqs. This limit corresponds to a column density of the $J=3$ level of $2 \times 10^{16} \ \rm{cm}^{-2}$ (see Eq.~\ref{Eq:cdint}), approximately two orders of magnitude higher than the detection threshold reached in absorption.

Interestingly, the upper envelope of the emission follows a well-defined trend that reflects the vertical distribution of molecular gas in the solar neighborhood. Assuming that the local diffuse ISM can be described by two partially molecular components, a spherical component associated with the Local Bubble and a slab representing the more extended Galactic ISM, the maximum expected emission can be written as
\begin{equation} \label{Eq:max}
I_{\rm max}(|b|) = \frac{C_1}{\max\left(\sin|b|,\, r \right)} + C_2 \left( \frac{1}{\sin|b|} - \frac{1}{\max\left(\sin|b|,\, r\right)} \right).
\end{equation}
$C_1 = 2 \times 10^{-7}$ and $C_2 = 10^{-6}$ are the maximum emissions (in \ecqs) of the S(1) line perpendicular to the Galactic plane associated with the Local Bubble component and the extended Galactic ISM, respectively, each adjusted to match the data, while $r = 0.5$ represents the fixed ratio between the characteristic height of the molecular gas layer and the radius of the Local Bubble \citep[e.g.,][]{Dame2001,Lallement2014}. The resulting curve, shown as a dashed line in Fig.~\ref{fig:s1vsb}, closely traces the upper envelope of the observed intensities, providing further indication that the emission arises predominantly from diffuse molecular gas.

This envelope and the distribution of observed S(1) intensities are found to be symmetric with respect to $b=0$. There is, however, a statistically significant difference in the detection rate of the S(1) line between northern and southern lines of sight. Although the sample contains twice as many northern lines of sight as southern ones, the detection rate is a factor of two higher in the south, independently of Galactic latitude. This difference cannot be attributed to a bias in integration time, as the northern and southern subsamples have identical integration time distributions. The asymmetry therefore reflects a higher probability of intercepting molecular regions along southern lines of sight, while the intrinsic emission properties of the gas remain unchanged, consistent with the distribution of molecular structures at high Galactic latitude revealed by \citet{Rohser2016}.

\subsection{Excitation properties}

\begin{figure}
    \centering
    \includegraphics[width=1\linewidth,trim = 2.3cm 2.5cm 0.7cm 3.2cm]{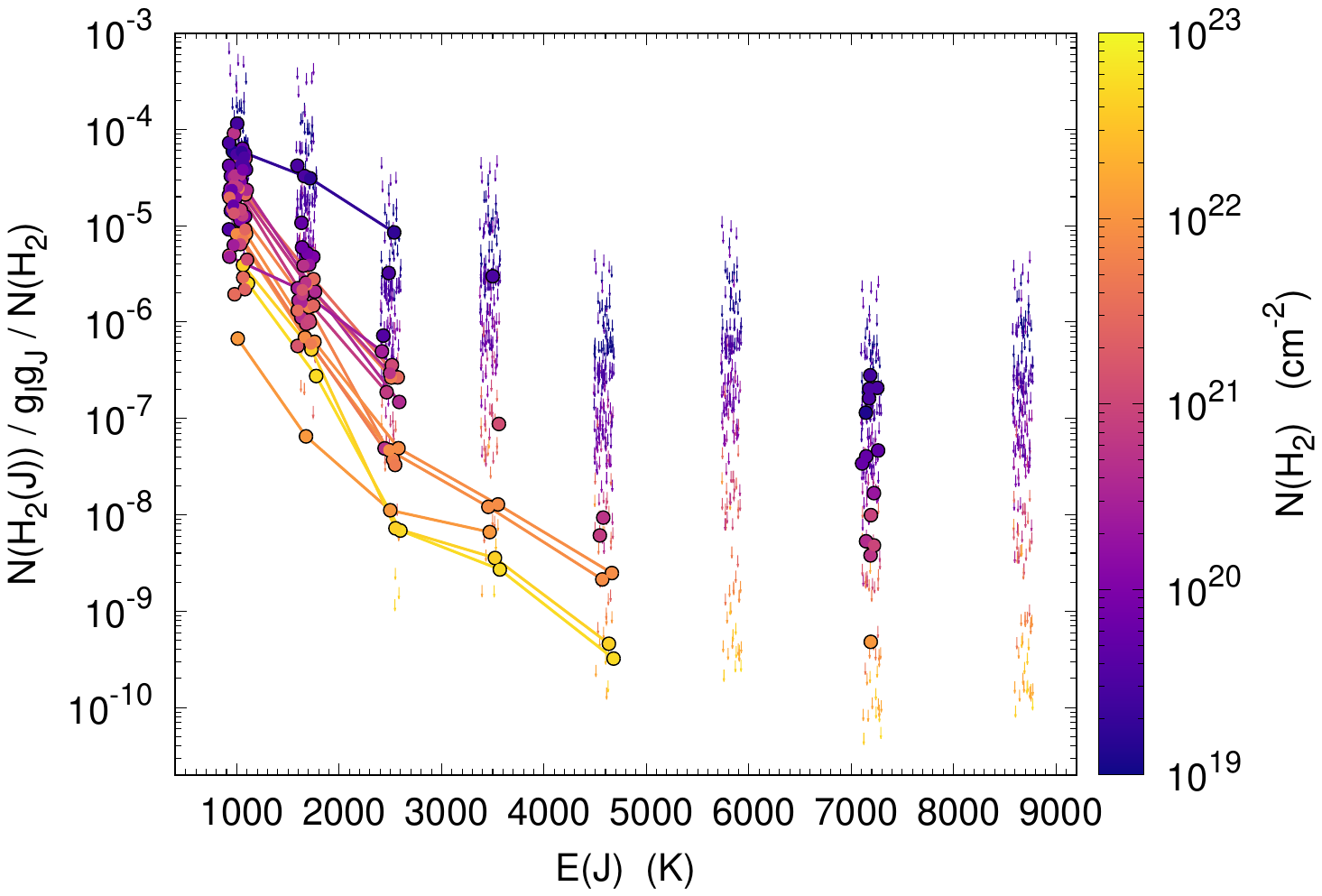}
    \caption{Excitation diagrams of \HH\ derived along the MRS background lines of sight, excluding the flagged data. The column densities of each rotational level are derived from the corresponding line intensities using Eq.~\ref{Eq:cdint} and are normalized and color-coded by the \HH\ column density estimated in Appendix \ref{append:ancillary}. For visual clarity, data points are slightly offset along the energy axis by small random values. Detections, along a given line of sight, of all successive rotational transitions up to at least the S(3) line are connected by segments.}
    \label{fig:excdiag}
\end{figure}

The distribution of \HH\ among its rotationally excited states is shown in Fig.~\ref{fig:excdiag}, which presents excitation diagrams constructed from the column densities of individual rotational levels, normalized by their statistical weights and by the estimated \HH\ column density (Appendix~\ref{append:ancillary}), as functions of the level energies. These excitation diagrams exhibit a striking consistency with those derived from absorption studies, in terms of both population ratios and the range of column densities probed (see Fig.~1.16 of \citealt{Godard2025a}). As found in previous work \citep[e.g.,][]{Gry2002}, the rotational populations depart from Local Thermal Equilibrium and display excitation temperatures that increase with the rotational quantum number $J$. In addition, and as reported by \citet{Wakker2006a}, the excitation of the rotational levels is linked to the column density of molecular hydrogen. As $N(\HH)$ increases, the fraction of \HH\ in its excited rotational states generally decreases, and the excitation diagrams presumably converge toward a mean distribution resulting from the statistical averaging along increasingly complex lines of sight.

\begin{figure}
    \centering
    \includegraphics[width=1\linewidth,trim = 0.0cm 1.0cm 0.0cm 0.5cm]{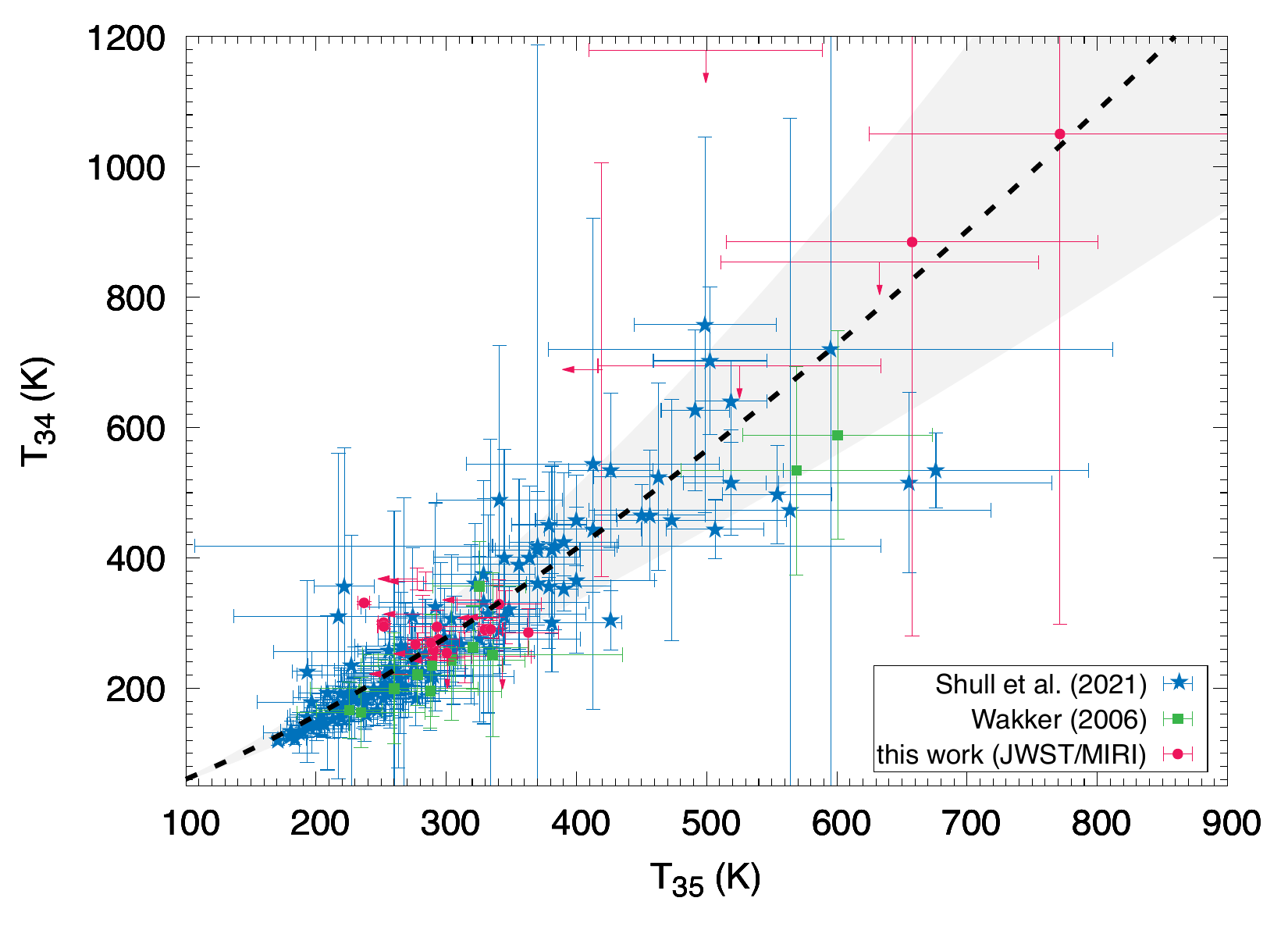}
    \caption{Excitation temperature $T_{34}$ between the $J=3$ and $J=4$ levels as a function of the excitation temperature $T_{35}$ between the $J=3$ and $J=5$ levels. Green squares and blue stars correspond to values inferred from absorption measurements toward extragalactic targets and nearby OB stars observed with FUSE \citep{Wakker2006a, Shull2021a}. Red points and arrows show measurements and upper limits, respectively, obtained along MRS background lines of sight after excluding flagged data. The dashed black line shows a power-law relation derived between the two temperatures (Eq.~\ref{eq:T34T35}), and the shaded region is an envelope obtained by varying each fit parameter independently within its marginalized $\pm 3\sigma$ uncertainty, which encloses 78\% of the data points.}
    \label{fig:T34vsT35}
\end{figure}

To further quantify the excitation of \HH\ in the diffuse ISM, we display in Fig.~\ref{fig:T34vsT35} the excitation temperatures $T_{34}$ and $T_{35}$, inferred from the populations of the $J=3$, 4, and 5 rotational levels. Because it involves both para- and ortho-rotational levels, $T_{34}$ serves as a proxy for the overall excitation of \HH\ that encodes the combined effects of ortho--para conversion and nonreactive excitation processes. Keeping this important aspect in mind, $T_{34}$ and $T_{35}$ are found to lie between 200 and $\sim$1000~K, in agreement with the values derived from absorption studies. As shown in Fig.~\ref{fig:T34vsT35}, the two excitation temperatures, derived from our sample and the lines of sight observed in absorption, follow a simple power-law relation,
\begin{equation} \label{eq:T34T35}
T_{34} = (106 \pm 10)\ \mathrm{K}\ \left(\frac{T_{35}}{150\ \mathrm{K}}\right)^{1.39 \pm 0.12},
\end{equation}
where the errors on each parameter correspond to the 3$\sigma$ marginalized uncertainties derived from the posterior distribution. Taken together, the results presented in Figs.~\ref{fig:excdiag} and \ref{fig:T34vsT35} provide paramount empirical constraints on the excitation conditions of molecular hydrogen in the diffuse ISM. In particular, they encode statistical information on the variations of the ortho-to-para ratio and on the distribution of densities, temperatures, and UV irradiation along a random set of lines of sight.

It has been proposed that the rotational excitation of \HH\ is enhanced by UV pumping due to the interaction with the background source along lines of sight targeting hot OB-type stars \citep[e.g.,][]{Shull2021a}. In contrast, the JWST observations analyzed here probe \HH\ emission along lines of sight that do not target specific background sources and are, a priori, not tied to the presence of nearby massive stars. The fact that the excitation properties, in terms of both excitation diagrams and $T_{34}$--$T_{35}$ distributions, are similar suggests either that the contribution of local stellar environments is difficult to disentangle from the cumulative excitation along the line of sight, or that their net impact on the averaged excitation properties is limited. A natural interpretation is that the observed excitation arises from a combination of collisional excitation and radiative pumping by the ambient interstellar radiation field integrated along extended path lengths, as found in recent numerical simulations of the diffuse multiphase ISM (see Fig. 1.20 of \citealt{Godard2025a}).

\subsection{Cooling rate}

\begin{figure*}
    \centering
    \includegraphics[width=17cm, trim = 2.0cm 2.5cm 0.5cm 2.0cm]{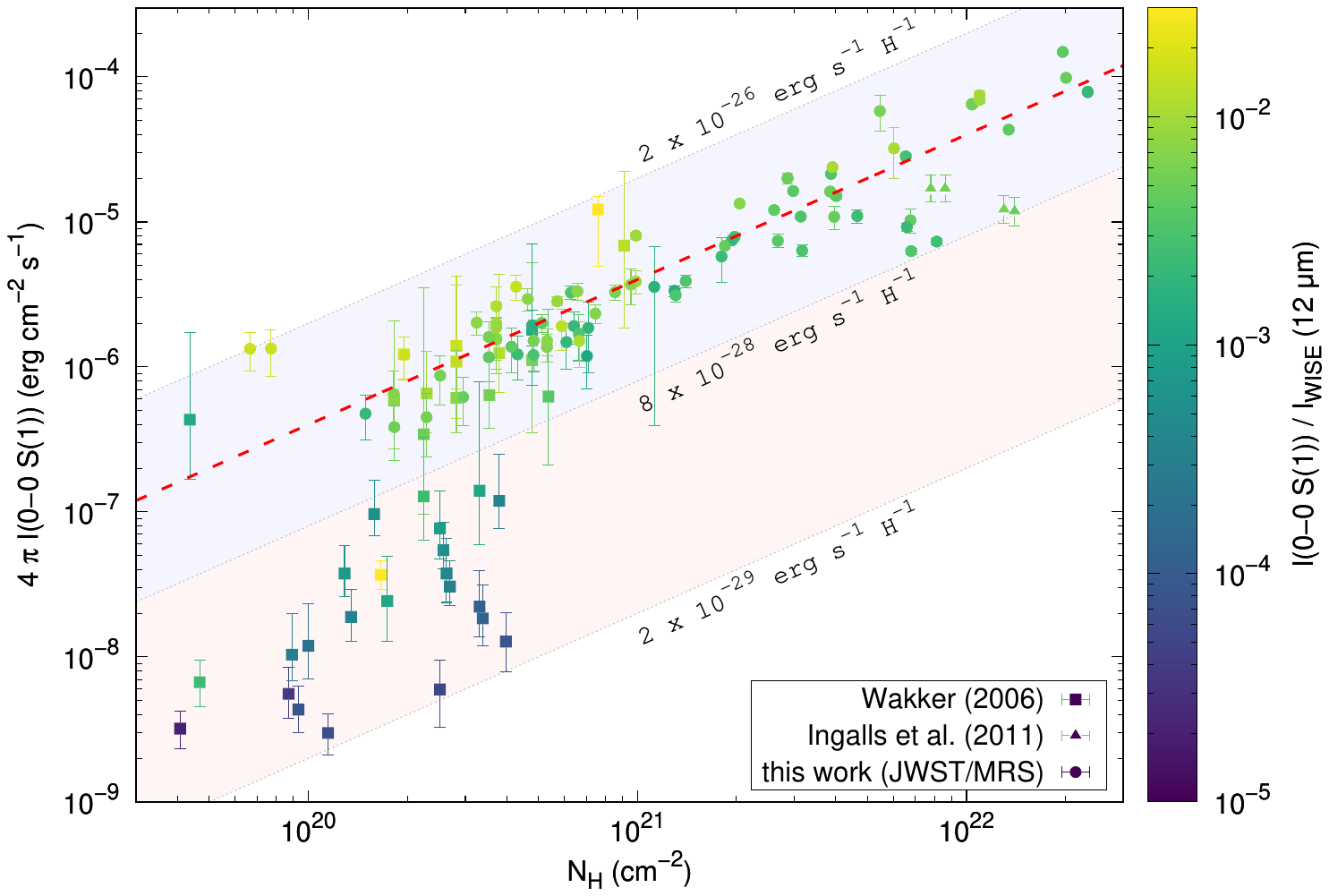}
    \caption{Flux of the 0--0 S(1) line of \HH\ as a function of the total hydrogen column density, $N_{\rm H}$ (Table~\ref{tab:ancillary}). Filled squares correspond to values inferred from absorption measurements toward extragalactic targets observed with FUSE \citep{Wakker2006a}. Filled circles show intensities measured along MRS background lines of sight after excluding flagged data. Points are color-coded by the ratio of the S(1) integrated intensity to the PAH surface brightness at 12~$\mu$m measured by WISE (Table~\ref{tab:ancillary}). The three dotted gray lines indicate constant values of the S(1) cooling rate per hydrogen atom. The dashed red line shows a fit to the S(1) cooling rate derived from the MRS sample only (Eq.~\ref{Eq:cool}).
    }
    \label{fig:S1vsNH}
\end{figure*}

The contribution of pure rotational \HH\ emission to the energy budget of the diffuse ISM is investigated in Fig.~\ref{fig:S1vsNH}, which shows the fluxes of the 0--0 S(1) line as a function of the total hydrogen column density, $N_{\rm H}$. Combining the MRS background measurements with the values inferred from absorption studies toward high-Galactic-latitude extragalactic sources reveals a sharp increase in S(1) emission at column densities between $10^{20}$ and $10^{21}$~cm$^{-2}$. This behavior coincides with the onset of the H--\HH\ transition in diffuse interstellar gas, marking the transition from a predominantly atomic to a partially molecular medium \citep[e.g.,][]{Bellomi2020}. Because emission measurements are less sensitive than absorption, the entire MRS sample probes lines of sight with substantial \HH\ column densities that have already undergone the H--\HH\ transition.

As established from absorption studies, the cooling associated with \HH\ emission per unit column density of gas, as well as its partition among the various rotational lines, is expected to vary from one line of sight to another. In particular, the 0--0 S(1) line typically contributes between 5\% and 95\% of the cooling carried by the rotational ladder. Nevertheless, Fig.~\ref{fig:S1vsNH} shows that once the H--\HH\ transition is reached, the cooling rate of the S(1) line per hydrogen atom,
\begin{equation} \label{Eq:cool}
\Lambda_{S(1)} = \frac{4\pi I(0-0\ S(1))}{N_{\rm H}} = 10^{-26.4\pm 0.3}\ {\rm erg}\ {\rm s}^{-1}\ {\rm H}^{-1}
\end{equation}
remains nearly constant, with a dispersion of less than a factor of two over nearly three orders of magnitude in total hydrogen column density. This cooling rate corresponds to about 20\% of the cooling rate of the [C\,II] 158\,$\mu$m line measured in diffuse gas at high Galactic latitudes \citep{Bock1993,Lehner2004}. Over the same range of column density, the S(1) line intensity amounts to $0.6\%$ of the polycyclic aromatic hydrocarbon (PAH) surface brightness at 12~$\mu$m measured by the Wide-field Infrared Survey Explorer (WISE) \citep{Meisner2014}, again with a dispersion smaller than a factor of two. These results are consistent with \citet{Hensley2016}, who showed that the PAH fraction increases in diffuse gas and reaches a plateau at $N_{\rm H} \gtrsim 8 \times 10^{20}$~cm$^{-2}$.

The fact that $\Lambda_{S(1)}$ and its ratio to the 12~$\mu$m PAH emission remain approximately constant beyond the H--\HH\ transition constitutes a new and important result. It provides a link between the physics of WNM--CNM phase transitions, CNM irradiation, \HH\ formation, ortho-to-para conversion processes, and \HH\ excitation in diffuse gas. In particular, the value inferred from the MRS data (Eq.~\ref{Eq:cool}) is larger than that predicted by models of static photodissociation regions as recently found by \citet{Goldsmith2025}.

\section{Conclusions and perspectives}
\label{Conclusion}

This paper introduces a novel observational strategy for probing the diffuse ISM that requires no dedicated telescope time. It demonstrates that diffuse \HH\ emission can be readily detected by systematically analyzing the backgrounds of existing JWST observations. This method significantly enhances the scientific yield of the JWST by leveraging data that would otherwise remain unused. The main results can be summarized as follows.
\begin{itemize}
    \item Of the 297 background observations acquired over the past three years with the MRS instrument, 233 are identified as probing the diffuse, predominantly local ISM. We report the detection of the pure rotational S(1) line of \HH\ along 41\% of these lines of sight, together with the first detections of the S(4), S(5), and S(7) pure rotational lines in the diffuse ISM. 
    \item Because of sensitivity limits, most detections correspond to lines of sight that have undergone the H--\HH\ transition, with molecular hydrogen column densities between $10^{19}$ and $10^{22}$~cm$^{-2}$.
    \item The S(1) line is detected over a wide range of Galactic latitudes, up to $|b| = 78^\circ$. Its intensity spans more than two orders of magnitude and exhibits a steep decline with increasing Galactic latitude, reflecting both the finite vertical extent of molecular gas above the Galactic disk and the radial extent of the Local Bubble.
    \item The intrinsic emission properties of \HH\ are symmetric with respect to $b=0$. However, the detection rate differs between northern and southern Galactic lines of sight, with the S(1) line detected twice as often in the south.
    \item The spatial distribution and excitation properties of \HH\ emission are consistent with those inferred from absorption studies conducted with FUSE. In particular, the excitation temperatures $T_{34}$ and $T_{35}$, which trace ortho--para conversion and nonreactive excitation processes, lie between 200 and $\sim$1000~K. These temperatures are well correlated with each other and anticorrelated with the \HH\ column density.
    \item Beyond expanding the statistical sample of Galactic H$_2$ emission, this study reveals a new cooling law for the diffuse ISM. When contextualized with \textit{Planck} and WISE data, lines of sight that have undergone the H--H$_2$ transition show a remarkably constant S(1) cooling rate per hydrogen atom, with a mean value of $\sim 4 \times 10^{-27}$~erg~s$^{-1}$~H$^{-1}$ and a dispersion of less than a factor of two over nearly three orders of magnitude in total column density. The S(1) line intensity amounts to 20\% of the cooling rate of the [C\,II] 158\,$\mu$m line derived at high Galactic latitudes, and 0.6\%  of the PAH surface brightness at 12$\mu$m, again with a dispersion smaller than a factor of two.
\end{itemize}

The widespread emission from rotationally excited \HH\ across the Galaxy offers a powerful complement to traditional absorption studies. While absorption data provide precise measurements of total \HH\ column densities and the ortho-to-para ratio with unmatched sensitivity, they are limited to rare, bright UV background sources. In contrast, current and future JWST observations enable a dramatic increase in statistical coverage across the sky and access to higher rotational levels that were previously undetectable in the diffuse ISM. In this study we focused on MRS data to demonstrate that this emission arises from diffuse gas. Ongoing work focuses on NIRSpec data, opening the door to the first systematic detection of diffuse rovibrational \HH\ emission.

\section*{Data availability}

The full versions of Tables~\ref{tab:all-detect} and~\ref{tab:ancillary} are only available in electronic form at the CDS via anonymous ftp to \href{http://cdsarc.u-strasbg.fr}{cdsarc.u-strasbg.fr} 
(130.79.128.5) or at \url{http://cdsweb.u-strasbg.fr/cgi-bin/qcat?J/A+A/}.

\begin{acknowledgements}
We thank the referee, J. M. Shull, for a thorough reading of the manuscript and for comments that greatly improved the paper. This work is based on observations made with the NASA/ESA/CSA James Webb Space Telescope. The data were obtained from the Mikulski Archive for Space Telescopes at the Space Telescope Science Institute, which is operated by the Association of Universities for Research in Astronomy, Inc., under NASA contract NAS 5-03127 for JWST. The research leading to these results has received financial support by the DIM ORIGINES program from Région Île de France. We would also like to acknowledge the support from the Thematic Actions "Cosmologie et Galaxies" (CG) and "Physique et Chimie du Milieu Interstellaire" (PCMI) of INSU Programme National "Astro", with contributions from CNRS Physique and CNRS Chimie, INP and IN2P3, CEA, and CNES. The data gathering and processing was made possible by using the CANDIDE cluster at the Institut d’Astrophysique de Paris. The cluster was funded through grants from the PNCG, CNES, DIM-ACAV, the Euclid Consortium, and the Danish National Research Foundation Cosmic Dawn Center (DNRF140). It is maintained by Stephane Rouberol.
\end{acknowledgements}

\bibliographystyle{aa} 
\bibliography{final_bib}

\begin{thebibliography}{56}
\expandafter\ifx\csname natexlab\endcsname\relax\def\natexlab#1{#1}\fi

\bibitem[{{{\'A}lvarez-M{\'a}rquez} {et~al.}(2023){{\'A}lvarez-M{\'a}rquez}, {Labiano}, {Guillard}, {Dicken}, {Argyriou}, {Patapis}, {Law}, {Kavanagh}, {Larson}, {Gasman}, {Mueller}, {Alberts}, {Brandl}, {Colina}, {Garc{\'\i}a-Mar{\'\i}n}, {Jones}, {Noriega-Crespo}, {Shivaei}, {Temim}, \& {Wright}}]{Alvarez2023}
{{\'A}lvarez-M{\'a}rquez}, J., {Labiano}, A., {Guillard}, P., {et~al.} 2023, \aap, 672, A108

\bibitem[{{Argyriou} {et~al.}(2023){Argyriou}, {Glasse}, {Law}, {Labiano}, {{\'A}lvarez-M{\'a}rquez}, {Patapis}, {Kavanagh}, {Gasman}, {Mueller}, {Larson}, {Vandenbussche}, {Glauser}, {Royer}, {Dicken}, {Harkett}, {Sargent}, {Engesser}, {Jones}, {Kendrew}, {Noriega-Crespo}, {Brandl}, {Rieke}, {Wright}, {Lee}, \& {Wells}}]{Argyriou2023}
{Argyriou}, I., {Glasse}, A., {Law}, D.~R., {et~al.} 2023, \aap, 675, A111

\bibitem[{{Bellomi} {et~al.}(2020){Bellomi}, {Godard}, {Hennebelle}, {Valdivia}, {Pineau des For{\^e}ts}, {Lesaffre}, \& {P{\'e}rault}}]{Bellomi2020}
{Bellomi}, E., {Godard}, B., {Hennebelle}, P., {et~al.} 2020, \aap, 643, A36

\bibitem[{{Bock} {et~al.}(1993){Bock}, {Hristov}, {Kawada}, {Matsuhara}, {Matsumoto}, {Matsuura}, {Mauskopf}, {Richards}, {Tanaka}, \& {Lange}}]{Bock1993}
{Bock}, J.~J., {Hristov}, V.~V., {Kawada}, M., {et~al.} 1993, \apjl, 410, L115

\bibitem[{{Bockel{\'e}e-Morvan} {et~al.}(2024){Bockel{\'e}e-Morvan}, {Lellouch}, {Poch}, {Quirico}, {Cazaux}, {de Pater}, {Fouchet}, {Fry}, {Rodriguez-Ovalle}, {Tosi}, {Wong}, {Boshuizen}, {de Kleer}, {Fletcher}, {Meunier}, {Mura}, {Roth}, {Saur}, {Schmitt}, {Trumbo}, {Brown}, {O'Donoghue}, {Orton}, \& {Showalter}}]{Bockelee-Morvan2024}
{Bockel{\'e}e-Morvan}, D., {Lellouch}, E., {Poch}, O., {et~al.} 2024, \aap, 681, A27

\bibitem[{{Bushouse} {et~al.}(2025){Bushouse}, {Eisenhamer}, {Dencheva}, {Davies}, {Greenfield}, {Morrison}, {Hodge}, {Simon}, {Grumm}, {Droettboom}, {Slavich}, {Sosey}, {Pauly}, {Miller}, {Jedrzejewski}, {Hack}, {Davis}, {Crawford}, {Law}, {Gordon}, {Regan}, {Cara}, {MacDonald}, {Bradley}, {Shanahan}, {Jamieson}, {Teodoro}, {Williams}, {Pena-Guerrero}, {Graham}, {Molter}, {Brandt}, {Hayes}, {Cooper}, {Clarke}, \& {Filippazzo}}]{Bushouse_1.17.1}
{Bushouse}, H., {Eisenhamer}, J., {Dencheva}, N., {et~al.} 2025, {JWST Calibration Pipeline}

\bibitem[{{Carruthers}(1970)}]{Carruthers1970}
{Carruthers}, G.~R. 1970, \apjl, 161, L81

\bibitem[{{Castro Neto} {et~al.}(2009){Castro Neto}, {Guinea}, {Peres}, {Novoselov}, \& {Geim}}]{CastroNeto2009}
{Castro Neto}, A.~H., {Guinea}, F., {Peres}, N.~M.~R., {Novoselov}, K.~S., \& {Geim}, A.~K. 2009, Reviews of Modern Physics, 81, 109

\bibitem[{{Compi{\`e}gne} {et~al.}(2011){Compi{\`e}gne}, {Verstraete}, {Jones}, {Bernard}, {Boulanger}, {Flagey}, {Le Bourlot}, {Paradis}, \& {Ysard}}]{Compiegne2011}
{Compi{\`e}gne}, M., {Verstraete}, L., {Jones}, A., {et~al.} 2011, \aap, 525, A103

\bibitem[{{Dame} {et~al.}(2001){Dame}, {Hartmann}, \& {Thaddeus}}]{Dame2001}
{Dame}, T.~M., {Hartmann}, D., \& {Thaddeus}, P. 2001, \apj, 547, 792

\bibitem[{{Draine} \& {Li}(2007)}]{Draine2007}
{Draine}, B.~T. \& {Li}, A. 2007, \apj, 657, 810

\bibitem[{{Falgarone} {et~al.}(2005){Falgarone}, {Verstraete}, {Pineau des For{\^e}ts}, \& {Hily-Blant}}]{Falgarone2005}
{Falgarone}, E., {Verstraete}, L., {Pineau des For{\^e}ts}, G., \& {Hily-Blant}, P. 2005, \aap, 433, 997

\bibitem[{{Frisch} \& {Jura}(1980)}]{Frisch1980}
{Frisch}, P.~C. \& {Jura}, M. 1980, \apj, 242, 560

\bibitem[{{Gillmon} \& {Shull}(2006)}]{Gillmon2006b}
{Gillmon}, K. \& {Shull}, J.~M. 2006, \apj, 636, 908

\bibitem[{{Gillmon} {et~al.}(2006){Gillmon}, {Shull}, {Tumlinson}, \& {Danforth}}]{Gillmon2006}
{Gillmon}, K., {Shull}, J.~M., {Tumlinson}, J., \& {Danforth}, C. 2006, \apj, 636, 891

\bibitem[{{Godard}(2025)}]{Godard2025a}
{Godard}, B. 2025, Habilitation {\`a} diriger des recherches, {Observatoire de Paris}

\bibitem[{{Godard} {et~al.}(2009){Godard}, {Falgarone}, \& {Pineau des For{\^e}ts}}]{Godard2009}
{Godard}, B., {Falgarone}, E., \& {Pineau des For{\^e}ts}, G. 2009, \aap, 495, 847

\bibitem[{{Godard} {et~al.}(2014){Godard}, {Falgarone}, \& {Pineau des For{\^e}ts}}]{Godard2014}
{Godard}, B., {Falgarone}, E., \& {Pineau des For{\^e}ts}, G. 2014, \aap, 570, A27

\bibitem[{{Godard} {et~al.}(2023){Godard}, {Pineau des For{\^e}ts}, {Hennebelle}, {Bellomi}, \& {Valdivia}}]{Godard2023}
{Godard}, B., {Pineau des For{\^e}ts}, G., {Hennebelle}, P., {Bellomi}, E., \& {Valdivia}, V. 2023, \aap, 669, A74

\bibitem[{{Goldsmith} {et~al.}(2025){Goldsmith}, {Wang}, {Wang}, {Skalidis}, {Fuller}, {Li}, {Tsai}, {Wang}, \& {Quan}}]{Goldsmith2025}
{Goldsmith}, P.~F., {Wang}, S., {Wang}, X., {et~al.} 2025, \apjl, 985, L4

\bibitem[{Gry {et~al.}(2002)Gry, Boulanger, Nehmé, Pineau~des Forêts, Habart, \& Falgarone}]{Gry2002}
Gry, C., Boulanger, F., Nehmé, C., {et~al.} 2002, \aap, 391, 675

\bibitem[{{Hensley} {et~al.}(2016){Hensley}, {Draine}, \& {Meisner}}]{Hensley2016}
{Hensley}, B.~S., {Draine}, B.~T., \& {Meisner}, A.~M. 2016, \apj, 827, 45

\bibitem[{{HI4PI Collaboration} {et~al.}(2016){HI4PI Collaboration}, {Ben Bekhti}, {Fl{\"o}er}, {Keller}, {Kerp}, {Lenz}, {Winkel}, {Bailin}, {Calabretta}, {Dedes}, {Ford}, {Gibson}, {Haud}, {Janowiecki}, {Kalberla}, {Lockman}, {McClure-Griffiths}, {Murphy}, {Nakanishi}, {Pisano}, \& {Staveley-Smith}}]{HI4PI2016}
{HI4PI Collaboration}, {Ben Bekhti}, N., {Fl{\"o}er}, L., {et~al.} 2016, \aap, 594, A116

\bibitem[{{Ingalls} {et~al.}(2011){Ingalls}, {Bania}, {Boulanger}, {Draine}, {Falgarone}, \& {Hily-Blant}}]{Ingalls2011a}
{Ingalls}, J.~G., {Bania}, T.~M., {Boulanger}, F., {et~al.} 2011, \apj, 743, 174

\bibitem[{{Labiano} {et~al.}(2021){Labiano}, {Argyriou}, {{\'A}lvarez-M{\'a}rquez}, {Glasse}, {Glauser}, {Patapis}, {Law}, {Brandl}, {Justtanont}, {Lahuis}, {Mart{\'\i}nez-Galarza}, {Mueller}, {Noriega-Crespo}, {Royer}, {Shaughnessy}, \& {Vandenbussche}}]{Labiano2021}
{Labiano}, A., {Argyriou}, I., {{\'A}lvarez-M{\'a}rquez}, J., {et~al.} 2021, \aap, 656, A57

\bibitem[{Labiano {et~al.}(2016)Labiano, Azzollini, Bailey, Beard, Dicken, García-Marín, Geers, Glasse, Glauser, Gordon, Justtanont, Klaassen, Lahuis, Law, Morrison, Müller, Rieke, Vandenbussche, \& Wright}]{labiano_miri_2016}
Labiano, A., Azzollini, R., Bailey, J., {et~al.} 2016, in Observatory {Operations}: {Strategies}, {Processes}, and {Systems} {VI}, Vol. 9910 (SPIE), 947--956

\bibitem[{{Lallement} {et~al.}(2014){Lallement}, {Vergely}, {Valette}, {Puspitarini}, {Eyer}, \& {Casagrande}}]{Lallement2014}
{Lallement}, R., {Vergely}, J.-L., {Valette}, B., {et~al.} 2014, \aap, 561, A91

\bibitem[{{Lambert} \& {Danks}(1986)}]{Lambert1986}
{Lambert}, D.~L. \& {Danks}, A.~C. 1986, \apj, 303, 401

\bibitem[{{Law} {et~al.}(2025){Law}, {Argyriou}, {Gordon}, {Sloan}, {Gasman}, {Glasse}, {Larson}, {Fletcher}, {Labiano}, \& {Noriega-Crespo}}]{Law2025}
{Law}, D.~R., {Argyriou}, I., {Gordon}, K.~D., {et~al.} 2025, \aj, 169, 67

\bibitem[{{Law} {et~al.}(2023){Law}, {E. Morrison}, {Argyriou}, {Patapis}, {{\'A}lvarez-M{\'a}rquez}, {Labiano}, \& {Vandenbussche}}]{Law2023}
{Law}, D.~R., {E. Morrison}, J., {Argyriou}, I., {et~al.} 2023, \aj, 166, 45

\bibitem[{{Lehner} {et~al.}(2004){Lehner}, {Wakker}, \& {Savage}}]{Lehner2004}
{Lehner}, N., {Wakker}, B.~P., \& {Savage}, B.~D. 2004, \apj, 615, 767

\bibitem[{{Leroy} {et~al.}(2013){Leroy}, {Walter}, {Sandstrom}, {Schruba}, {Munoz-Mateos}, {Bigiel}, {Bolatto}, {Brinks}, {de Blok}, {Meidt}, {Rix}, {Rosolowsky}, {Schinnerer}, {Schuster}, \& {Usero}}]{Leroy2013a}
{Leroy}, A.~K., {Walter}, F., {Sandstrom}, K., {et~al.} 2013, \aj, 146, 19

\bibitem[{{Lesaffre} {et~al.}(2020){Lesaffre}, {Todorov}, {Levrier}, {Valdivia}, {Dzyurkevich}, {Godard}, {Tram}, {Gusdorf}, {Lehmann}, \& {Falgarone}}]{Lesaffre2020}
{Lesaffre}, P., {Todorov}, P., {Levrier}, F., {et~al.} 2020, \mnras, 495, 816

\bibitem[{{Lin} {et~al.}(2019){Lin}, {Pan}, {Ellison}, {Belfiore}, {Shi}, {S{\'a}nchez}, {Hsieh}, {Rowlands}, {Ramya}, {Thorp}, {Li}, \& {Maiolino}}]{Lin2019a}
{Lin}, L., {Pan}, H.-A., {Ellison}, S.~L., {et~al.} 2019, \apjl, 884, L33

\bibitem[{{Meisner} \& {Finkbeiner}(2014)}]{Meisner2014}
{Meisner}, A.~M. \& {Finkbeiner}, D.~P. 2014, \apj, 781, 5

\bibitem[{{Miville-Desch{\^e}nes} {et~al.}(2017){Miville-Desch{\^e}nes}, {Murray}, \& {Lee}}]{Miville-Deschenes2017}
{Miville-Desch{\^e}nes}, M.-A., {Murray}, N., \& {Lee}, E.~J. 2017, \apj, 834, 57

\bibitem[{{Morrison} {et~al.}(2023){Morrison}, {Dicken}, {Argyriou}, {Ressler}, {Gordon}, {Regan}, {Cracraft}, {Rieke}, {Engesser}, {Alberts}, {Alvarez-Marquez}, {Colbert}, {Fox}, {Gasman}, {Law}, {Garcia Marin}, {G{\'a}sp{\'a}r}, {Guillard}, {Kendrew}, {Labiano}, {Laine}, {Noriega-Crespo}, {Shivaei}, \& {Sloan}}]{Morrison2023}
{Morrison}, J.~E., {Dicken}, D., {Argyriou}, I., {et~al.} 2023, \pasp, 135, 075004

\bibitem[{{Moseley} {et~al.}(2021){Moseley}, {Draine}, {Tomida}, \& {Stone}}]{Moseley2021}
{Moseley}, E.~R., {Draine}, B.~T., {Tomida}, K., \& {Stone}, J.~M. 2021, \mnras, 500, 3290

\bibitem[{{Myers} {et~al.}(2015){Myers}, {McKee}, \& {Li}}]{Myers2015}
{Myers}, A.~T., {McKee}, C.~F., \& {Li}, P.~S. 2015, \mnras, 453, 2747

\bibitem[{{Nehm{\'e}} {et~al.}(2008){Nehm{\'e}}, {Le Bourlot}, {Boulanger}, {Pineau des For{\^e}ts}, \& {Gry}}]{Nehme2008}
{Nehm{\'e}}, C., {Le Bourlot}, J., {Boulanger}, F., {Pineau des For{\^e}ts}, G., \& {Gry}, C. 2008, \aap, 483, 485

\bibitem[{{Pessa} {et~al.}(2022){Pessa}, {Schinnerer}, {Leroy}, {Koch}, {Rosolowsky}, {Williams}, {Pan}, {Schruba}, {Usero}, {Belfiore}, {Bigiel}, {Blanc}, {Chevance}, {Dale}, {Emsellem}, {Gensior}, {Glover}, {Grasha}, {Groves}, {Klessen}, {Kreckel}, {Kruijssen}, {Liu}, {Meidt}, {Pety}, {Querejeta}, {Saito}, {Sanchez-Blazquez}, \& {Watkins}}]{Pessa2022a}
{Pessa}, I., {Schinnerer}, E., {Leroy}, A.~K., {et~al.} 2022, \aap, 663, A61

\bibitem[{{Planck Collaboration} {et~al.}(2014){Planck Collaboration}, {Abergel}, {Ade}, {Aghanim}, {Alves}, {Aniano}, {Armitage-Caplan}, {Arnaud}, {Ashdown}, {Atrio-Barandela}, {Aumont}, {Baccigalupi}, {Banday}, {Barreiro}, {Bartlett}, {Battaner}, {Benabed}, {Beno{\^\i}t}, {Benoit-L{\'e}vy}, {Bernard}, {Bersanelli}, {Bielewicz}, {Bobin}, {Bock}, {Bonaldi}, {Bond}, {Borrill}, {Bouchet}, {Boulanger}, {Bridges}, {Bucher}, {Burigana}, {Butler}, {Cardoso}, {Catalano}, {Chamballu}, {Chary}, {Chiang}, {Chiang}, {Christensen}, {Church}, {Clemens}, {Clements}, {Colombi}, {Colombo}, {Combet}, {Couchot}, {Coulais}, {Crill}, {Curto}, {Cuttaia}, {Danese}, {Davies}, {Davis}, {de Bernardis}, {de Rosa}, {de Zotti}, {Delabrouille}, {Delouis}, {D{\'e}sert}, {Dickinson}, {Diego}, {Dole}, {Donzelli}, {Dor{\'e}}, {Douspis}, {Draine}, {Dupac}, {Efstathiou}, {En{\ss}lin}, {Eriksen}, {Falgarone}, {Finelli}, {Forni}, {Frailis}, {Fraisse}, {Franceschi}, {Galeotta}, {Ganga}, {Ghosh}, {Giard}, {Giardino}, {Giraud-H{\'e}raud},
  {Gonz{\'a}lez-Nuevo}, {G{\'o}rski}, {Gratton}, {Gregorio}, {Grenier}, {Gruppuso}, {Guillet}, {Hansen}, {Hanson}, {Harrison}, {Helou}, {Henrot-Versill{\'e}}, {Hern{\'a}ndez-Monteagudo}, {Herranz}, {Hildebrandt}, {Hivon}, {Hobson}, {Holmes}, {Hornstrup}, {Hovest}, {Huffenberger}, {Jaffe}, {Jaffe}, {Jewell}, {Joncas}, {Jones}, {Juvela}, {Keih{\"a}nen}, {Keskitalo}, {Kisner}, {Knoche}, {Knox}, {Kunz}, {Kurki-Suonio}, {Lagache}, {L{\"a}hteenm{\"a}ki}, {Lamarre}, {Lasenby}, {Laureijs}, {Lawrence}, {Leonardi}, {Le{\'o}n-Tavares}, {Lesgourgues}, {Levrier}, {Liguori}, {Lilje}, {Linden-V{\o}rnle}, {L{\'o}pez-Caniego}, {Lubin}, {Mac{\'\i}as-P{\'e}rez}, {Maffei}, {Maino}, {Mandolesi}, {Maris}, {Marshall}, {Martin}, {Mart{\'\i}nez-Gonz{\'a}lez}, {Masi}, {Massardi}, {Matarrese}, {Matthai}, {Mazzotta}, {McGehee}, {Melchiorri}, {Mendes}, {Mennella}, {Migliaccio}, {Mitra}, {Miville-Desch{\^e}nes}, {Moneti}, {Montier}, {Morgante}, {Mortlock}, {Munshi}, {Murphy}, {Naselsky}, {Nati}, {Natoli}, {Netterfield},
  {N{\o}rgaard-Nielsen}, {Noviello}, {Novikov}, {Novikov}, {Osborne}, {Oxborrow}, {Paci}, {Pagano}, {Pajot}, {Paladini}, {Paoletti}, {Pasian}, {Patanchon}, {Perdereau}, {Perotto}, {Perrotta}, {Piacentini}, {Piat}, {Pierpaoli}, {Pietrobon}, {Plaszczynski}, {Pointecouteau}, {Polenta}, {Ponthieu}, {Popa}, {Poutanen}, {Pratt}, {Pr{\'e}zeau}, {Prunet}, {Puget}, {Rachen}, {Reach}, {Rebolo}, {Reinecke}, {Remazeilles}, {Renault}, {Ricciardi}, \& {Riller}}]{Planck2014}
{Planck Collaboration}, {Abergel}, A., {Ade}, P.~A.~R., {et~al.} 2014, \aap, 571, A11

\bibitem[{Rachford {et~al.}(2002)Rachford, Snow, Tumlinson, Shull, Blair, Ferlet, Friedman, Gry, Jenkins, Morton, Savage, Sonnentrucker, Vidal-Madjar, Welty, \& York}]{Rachford2002}
Rachford, B., Snow, T., Tumlinson, J., {et~al.} 2002, \apj, 577, 221

\bibitem[{{Rigby} {et~al.}(2023){Rigby}, {Lightsey}, {Garc{\'\i}a Mar{\'\i}n}, {Bowers}, {Smith}, {Glasse}, {McElwain}, {Rieke}, {Chary}, {Liu}, {Clampin}, {Kimble}, {Kinzel}, {Laidler}, {Mehalick}, {Noriega-Crespo}, {Shivaei}, {Skelton}, {Stark}, {Temim}, {Wei}, \& {Willott}}]{Rigby2023}
{Rigby}, J.~R., {Lightsey}, P.~A., {Garc{\'\i}a Mar{\'\i}n}, M., {et~al.} 2023, \pasp, 135, 048002

\bibitem[{{R{\"o}hser} {et~al.}(2016){R{\"o}hser}, {Kerp}, {Lenz}, \& {Winkel}}]{Rohser2016}
{R{\"o}hser}, T., {Kerp}, J., {Lenz}, D., \& {Winkel}, B. 2016, \aap, 596, A94

\bibitem[{{Roueff} {et~al.}(2019){Roueff}, {Abgrall}, {Czachorowski}, {Pachucki}, {Puchalski}, \& {Komasa}}]{Roueff2019}
{Roueff}, E., {Abgrall}, H., {Czachorowski}, P., {et~al.} 2019, \aap, 630, A58

\bibitem[{{Savage} {et~al.}(1977){Savage}, {Bohlin}, {Drake}, \& {Budich}}]{Savage1977}
{Savage}, B.~D., {Bohlin}, R.~C., {Drake}, J.~F., \& {Budich}, W. 1977, \apj, 216, 291

\bibitem[{{Shull} \& {Beckwith}(1982)}]{Shull1982}
{Shull}, J.~M. \& {Beckwith}, S. 1982, \araa, 20, 163

\bibitem[{{Shull} {et~al.}(2021){Shull}, {Danforth}, \& {Anderson}}]{Shull2021a}
{Shull}, J.~M., {Danforth}, C.~W., \& {Anderson}, K.~L. 2021, \apj, 911, 55

\bibitem[{{Shull} \& {Hollenbach}(1978)}]{Shull1978}
{Shull}, J.~M. \& {Hollenbach}, D.~J. 1978, \apj, 220, 525

\bibitem[{{Skalidis} {et~al.}(2024){Skalidis}, {Goldsmith}, {Hopkins}, \& {Ponnada}}]{Skalidis2024}
{Skalidis}, R., {Goldsmith}, P.~F., {Hopkins}, P.~F., \& {Ponnada}, S.~B. 2024, \aap, 682, A161

\bibitem[{{Spitzer} {et~al.}(1974){Spitzer}, {Cochran}, \& {Hirshfeld}}]{Spitzer1974}
{Spitzer}, Jr., L., {Cochran}, W.~D., \& {Hirshfeld}, A. 1974, \apjs, 28, 373

\bibitem[{{Spitzer} \& {Jenkins}(1975)}]{Spitzer1975}
{Spitzer}, Jr., L. \& {Jenkins}, E.~B. 1975, \araa, 13, 133

\bibitem[{{Valdivia} {et~al.}(2016){Valdivia}, {Hennebelle}, {G{\'e}rin}, \& {Lesaffre}}]{Valdivia2016}
{Valdivia}, V., {Hennebelle}, P., {G{\'e}rin}, M., \& {Lesaffre}, P. 2016, \aap, 587, A76

\bibitem[{{Wakker}(2006)}]{Wakker2006a}
{Wakker}, B.~P. 2006, \apjs, 163, 282

\bibitem[{{Wolfire} {et~al.}(2003){Wolfire}, {McKee}, {Hollenbach}, \& {Tielens}}]{Wolfire2003}
{Wolfire}, M.~G., {McKee}, C.~F., {Hollenbach}, D., \& {Tielens}, A. 2003, \apj, 587, 278

\end{thebibliography}

\begin{appendix}

\section{Spectra and integrated intensities}
\label{append:data}

\begin{figure*}[!]
    \centering
    \includegraphics[width=17cm,trim = 0.5cm 0cm 0.5cm 0cm]{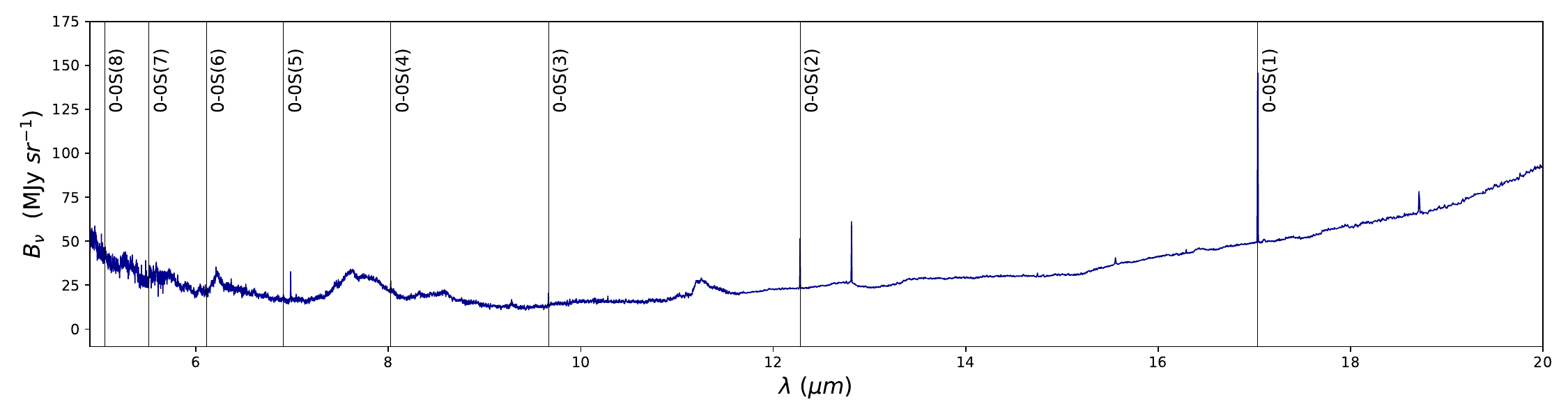}
    \includegraphics[width=17cm,trim = 0.5cm 0.5cm 0.5cm 0cm]{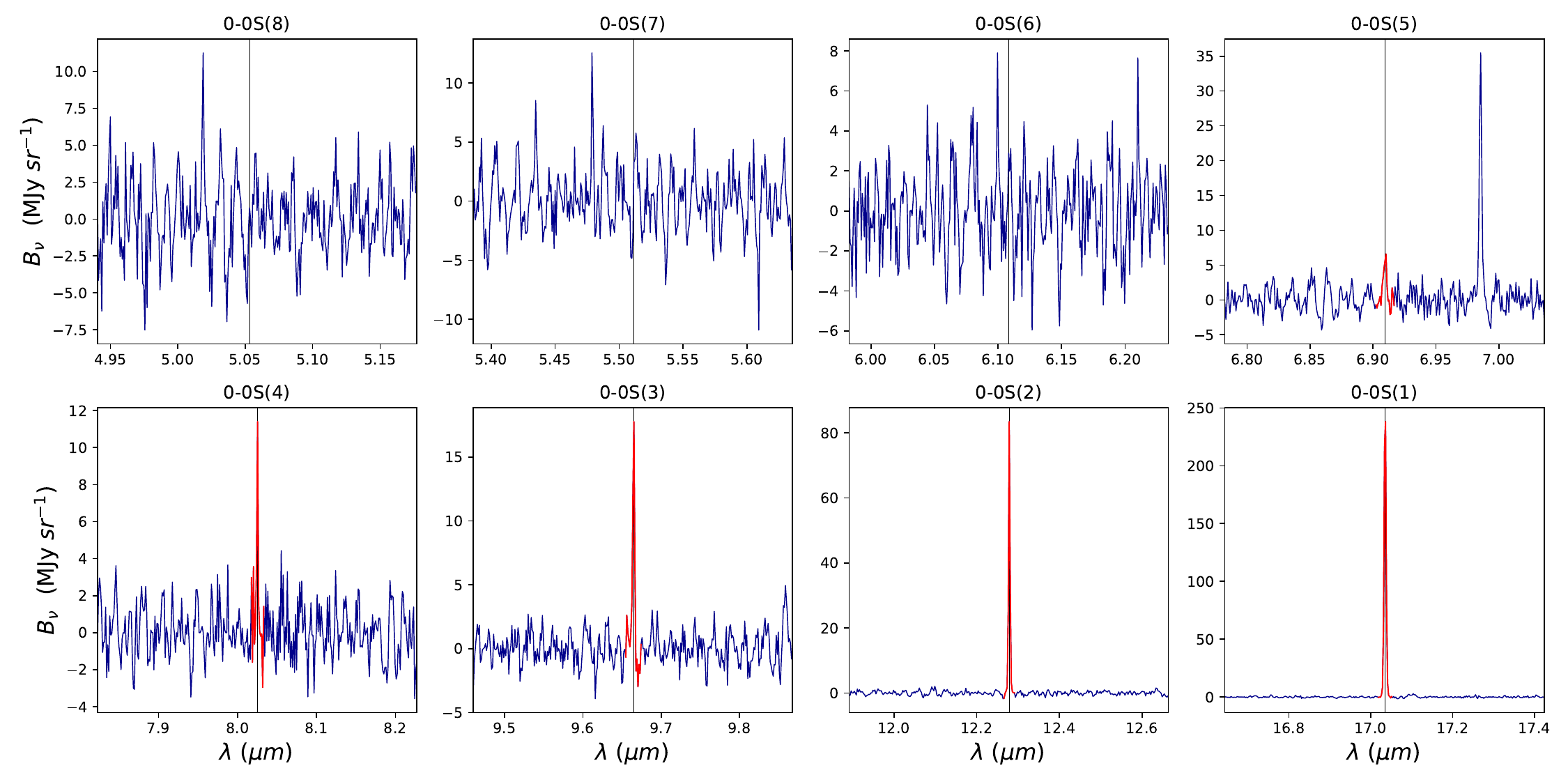}
    \caption{Example of a MRS background spectrum observed at Galactic coordinates $l,b=$ 12.881$^\circ$, 0.518$^\circ$ (Program ID: 01290, Observation ID: 078). {\it Top panel:} Observed specific intensity as a function of wavelength, after stitching the channel-specific spectra. {\it Bottom panels:} Continuum-subtracted intensity in zoomed-in windows centered on each \HH\ rotational line. Detected lines are highlighted in red.}
    \label{fig:spectrum}
\end{figure*}

\begin{table*}[!]
\centering
\caption{Subset of the full catalog of background observations retrieved from the \href{https://mast.stsci.edu/portal/Mashup/Clients/Mast/Portal.html}{MAST} archive and associated measured integrated intensities of \HH\ pure rotational lines.}
\label{tab:all-detect}
\begin{tabular}{rrrrrrrrrrrr}
\hline
PID & OID & flag & long       & lat        & S(1) & err & S(2) & err   & S(3) & err & ...  \\
    &     &      & ($^\circ$) & ($^\circ$) & \multicolumn{6}{c}{(\ecqs)} & ...  \\
\hline
01192 & 002 &    0 & 104.190 &  14.155 &  1.3 (-6) & 1.4 (-7) &  9.4 (-7) & 2.0 (-7) & <1.2 (-6) &          & ... \\
01192 & 011 &    0 & 206.846 & -16.758 &  9.2 (-7) & 6.4 (-8) &  4.9 (-7) & 2.0 (-7) & <5.3 (-7) &          & ... \\
01257 & 002 &    0 & 160.482 & -18.006 &  4.0 (-6) & 8.7 (-8) &  1.8 (-6) & 1.4 (-7) &  8.3 (-7) & 3.6 (-7) & ... \\
01267 & 004 &    0 &  36.646 &  53.026 & <7.5 (-8) &          & <1.2 (-7) &          & <2.8 (-7) &          & ... \\
...   & ... & ...  & ...     & ...     & ...       & ...      & ...       & ...      & ...       & ...      & ... \\
01023 & 038 &    1 & 271.870 & -32.823 &           &          & <9.3 (-6) &          &           &          & ... \\
01039 & 006 &    1 &  96.451 &  29.752 &           &          & <1.5 (-7) &          &           &          & ... \\
01049 & 004 &    1 & 280.831 & -32.783 & <6.0 (-8) &          & <1.3 (-7) &          & <2.2 (-7) &          & ... \\
01050 & 008 &    1 &  89.262 &  30.576 &  1.1 (-7) & 2.4 (-8) & <1.1 (-7) &          & <1.8 (-7) &          & ... \\
01050 & 010 &    1 &  89.262 &  30.576 &  1.2 (-7) & 2.6 (-8) &  1.6 (-7) & 6.2 (-8) & <2.0 (-7) &          & ... \\
01204 & 018 &    1 & 326.055 &  20.828 &           &          & <1.1 (-7) &          &           &          & ... \\
01204 & 020 &    1 &  22.219 &  19.352 &           &          &  1.5 (-7) & 6.0 (-8) &           &          & ... \\
01204 & 021 &    1 & 106.999 & -24.266 &           &          & <1.1 (-7) &          &           &          & ... \\
01253 & 013 &    1 & 194.390 & -28.832 & <1.3 (-7) &          & <2.7 (-7) &          & <6.0 (-7) &          & ... \\
01261 & 002 &    1 & 271.739 & -32.774 &           &          & <1.4 (-6) &          &           &          & ... \\
01265 & 002 &    1 &  20.751 &  27.278 &  9.7 (-8) & 3.2 (-8) & <1.8 (-7) &          & <2.7 (-7) &          & ... \\
01269 & 003 &    1 & 309.626 &  19.409 & <2.3 (-7) &          & <4.6 (-7) &          & <1.2 (-6) &          & ... \\
01282 & 019 &    1 & 278.683 &  22.963 & <2.1 (-7) &          & <4.9 (-7) &          & <9.6 (-7) &          & ... \\
...   & ... & ...  & ...     & ...     & ...       & ...      & ...       & ...      & ...       & ...      & ... \\
\hline
\end{tabular}
\tablefoot{
Program ID (PID) and Observation ID (OID) correspond to JWST identifiers. The flag column indicates the classification of each line of sight as diffuse (1) or non-diffuse (0) based on the selection criteria presented in Appendix~\ref{append:flag}. Numbers in parentheses are powers of ten. The full version of this table is available online at the CDS.\\
}
\end{table*}

\subsection{Spectral features}

As shown by \citet{Rigby2023}, MRS background spectra include several structured emission components: the thermal emission from the telescope, which dominates above 15~$\mu$m; the zodiacal light, which dominates below 15~$\mu$m; and the stray light, which becomes significant below 5~$\mu$m. We searched for \HH\ emission on top of this structured continuum and instrumental features, and above the instrument noise, which increases significantly near the edges of the spectral bands (below 7 and above 27~~$\mu$m).

As an illustration, we display in Fig.~\ref{fig:spectrum} a representative background spectrum observed in all four MRS IFU channels, along with zoomed-in panels centered on each \HH\ pure rotational line. The continuum emission is subtracted by applying a sliding median filter with sigma clipping, using a kernel width of 20 spectral resolution elements. Galactic \HH\ lines are search for by performing Gaussian fits over a velocity range of 600~\kms\ around the corresponding rest frequencies of each transition. The S(1) and S(2) lines are considered detected when their integrated intensities and maximum amplitude exceed the $3\sigma$ level, where the uncertainties $\sigma$ are estimated from the standard deviation of the continuum-subtracted spectrum. Higher-$J$ rotational lines are considered detected if their integrated intensities and maximum amplitude exceed $3\sigma$ and if at least one of the S(1) or S(2) lines is detected. Due to the strong instrumental noise beyond 27~$\mu$m \citep{Argyriou2023}, the 0-0 S(0) line of \HH\ is not included in our analysis.

\subsection{Line intensities}

A subset of the lines of sight retrieved from the \href{https://mast.stsci.edu/portal/Mashup/Clients/Mast/Portal.html}{MAST} archive, together with the integrated intensities of all detected \HH\ lines and their associated uncertainties, is presented in Table~\ref{tab:all-detect}. The full version of the table is available online at the Centre de Données astronomiques de Strasbourg (CDS). The sample covers a broad region of the sky, spanning all Galactic longitudes and latitudes (see Fig.~\ref{fig:aitoff}). Because the background observations originate from programs targeting sources with widely varying luminosities, the exposure times span a large range, from a few tens of seconds up to 46 minutes.

All pure rotational lines from S(1) to S(7) are detected at least once across the full observational sample. As described in Sect.~\ref{Sect:selection} and detailed in Appendix~\ref{append:flag}, selection criteria based on the total hydrogen column density and the average dust temperature are applied to identify lines of sight intersecting dense and strongly irradiated material possibly associated with star-forming complexes. When these flagged observations are excluded, the S(6) line is no longer detected.

\subsection{Detection statistics}

Not all lines of sight were observed in every MRS IFU channel. The dataset presented in Table~\ref{tab:all-detect} is therefore incomplete. The distribution of channel coverage across the sample is summarized in Table~\ref{tab:statistics}, which also provides detection statistics for each \HH\ pure rotational line. Restricting the analysis to diffuse lines of sight, the 0--0 S(1) line of \HH\ is detected along 84 lines of sight, corresponding to a detection rate of 41\%. The detection rate decreases steadily for higher-energy rotational transitions, with a few detections reported for the S(4), S(5), and S(7) lines. This trend reflects both the decreasing intrinsic brightness of higher-$J$ transitions and the limited sensitivity of the background observations.

\begin{table}
\centering
\caption{Summary of the 276 lines of sight analyzed in this paper (see Sect.~\ref{Sect:selection}) and corresponding numbers of detection of the pure rotational lines of \HH.}
\label{tab:statistics}
\begin{tabular}{l rr|rr|rr}
\hline
S($J$) & \multicolumn{1}{c}{$\lambda_0$} & \multicolumn{1}{c}{$A$}        & \multicolumn{2}{c}{total sample} & \multicolumn{2}{c}{detections} \\
       & \multicolumn{1}{c}{($\mu$m)}    & \multicolumn{1}{c}{(s$^{-1}$)} & all & diffuse                    & all & diffuse \\
\hline
S(0) & 28.2188 & 2.94 (-11) & --  & --      &  -- &     -- \\
S(1) & 17.0348 & 4.76 (-10) & 248 & 206     & 116 &     84 \\
S(2) & 12.2786 & 2.76 (-09) & 237 & 197     &  66 &     40 \\
S(3) &  9.6649 & 9.84 (-09) & 217 & 178     &  42 &     23 \\
S(4) &  8.0250 & 2.64 (-08) & 237 & 197     &  20 &      7 \\
S(5) &  6.9095 & 5.88 (-08) & 248 & 206     &  24 &      9 \\
S(6) &  6.1086 & 1.14 (-07) & 217 & 178     &   2 &      0 \\
S(7) &  5.5112 & 2.00 (-07) & 237 & 197     &  23 &     17 \\
S(8) &  5.0531 & 3.24 (-07) & 237 & 197     &   0 &      0 \\
 \hline
\end{tabular}
\tablefoot{Diffuse and non-diffuse lines of sight (flagged 1 and 0 in Table~\ref{tab:all-detect}, respectively) are identified based on the selection criteria presented in Appendix~\ref{append:flag}. The rest wavelengths, $\lambda_0$, and spontaneous-emission Einstein coefficients, $A$, of the pure rotational transitions of \HH\ are taken from \citet{Roueff2019}. Numbers in parenthesis are power of ten.
}
\end{table}

\subsection{Conversion to column densities}

Interpreting the observed \HH\ emission and performing comparisons with ancillary data derived from absorption studies requires converting integrated line intensities into column densities, and vice versa. Assuming optically thin emission, the integrated intensity of a transition, $I_{ul}$, can be converted into the column density of the upper level, $N_u$, according to
\begin{equation} \label{Eq:cdint}
N_u = \frac{4 \pi I_{ul}}{A_{ul} h \nu_{ul}},
\end{equation}
where $A_{ul}$ and $\nu_{ul}$ are the Einstein coefficient and frequency of the transition, respectively, and $h$ is the Planck constant. This approximation is valid provided that self-absorption by \HH\ and absorption by PAHs can be neglected.

Because most of the \HH\ column density resides in the $J=0$ and $J=1$ levels in the diffuse ISM, only the S(1) transition can potentially be affected by self-absorption among the pure rotational lines considered here. Using a lower-bound estimate for the line width of $1$~km~s$^{-1}$, the peak opacity of the S(1) line remains below 0.1 as long as $N(\HH) \lesssim 10^{23}$~cm$^{-2}$.

Absorption by PAHs can be estimated using conservative assumptions. Approximating PAHs as planar graphene flakes with a representative size of 6.4~\AA\ \citep{Compiegne2011} yields $\sim 50$ carbon atoms per PAH, based on the graphene lattice geometry \citep{CastroNeto2009}. For a PAH mass fraction relative to dust of 4.6\% in the diffuse ISM \citep{Draine2007} and a dust-to-gas mass ratio of 1\%, the PAH abundance relative to hydrogen is $x_{\rm PAH}\sim10^{-6}$. Using an upper-bound estimate of the absorption cross section per PAH over the MIRI wavelength range of $10^{-18}$~cm$^{2}$ \citep{Draine2007}, PAH absorption therefore remains negligible, with an optical depth below 0.1, as long as the total hydrogen column density $N_{\rm H} \lesssim 10^{23}$~cm$^{-2}$.

As shown in Appendix~\ref{append:ancillary} (Table~\ref{tab:ancillary}), both conditions, for negligible \HH\ self-absorption and negligible PAH absorption, are satisfied throughout our sample, except for the flagged lines of sight (see Sect.~\ref{Sect:selection} and Appendix~\ref{append:flag}). Throughout this paper, we therefore use intensity and upper-level column density interchangeably, enabling direct comparisons with absorption measurements and the construction of excitation diagrams (see Sect.~\ref{Sect:results}).

\section{Estimates of gas and dust properties}
\label{append:ancillary}

\begin{table*}[!]
\centering
\caption{Estimates of gas column densities and PAH and dust emission properties along the lines of sight listed in Table~\ref{tab:all-detect}.}
\label{tab:ancillary}
\begin{tabular}{rrrrrrrrrr}
\hline
PID & OID & flag & long       & lat        & T$_{\mathrm{dust}}$ & $N(\mathrm{H})$ & $N_{\mathrm{H}}$ & $\tau_{353}$      & $B_\nu$(12 $\mu$m) \\
    &     &      & ($^\circ$) & ($^\circ$) & (K)      & (cm$^{-2}$)     & (cm$^{-2}$)      &    & (MJy sr$^{-1}$)  \\
\hline
01192 & 002 &    0 & 104.190 &  14.155 &  19.36 & 2.2 (+21) & 1.1 (+22) & 1.3 (-4) &     2.219 \\ 
01192 & 011 &    0 & 206.846 & -16.758 &  27.43 & 2.0 (+21) & 6.6 (+21) & 7.9 (-5) &    10.948 \\ 
01257 & 002 &    0 & 160.482 & -18.006 &  17.16 & 1.2 (+21) & 5.5 (+22) & 6.6 (-4) &     4.988 \\ 
01267 & 004 &    0 &  36.646 &  53.026 &  27.92 & 4.0 (+20) & 5.0 (+20) & 6.0 (-6) &     0.080 \\ 
...   & ... & ...  & ...     & ...     & ...    &  ...      & ...       & ...      &       ... \\
01023 & 038 &    1 & 271.870 & -32.823 &  19.34 & 4.1 (+20) & 5.2 (+20) & 3.3 (-6) &    0.086  \\
01039 & 006 &    1 &  96.451 &  29.752 &  21.63 & 3.7 (+20) & 4.3 (+20) & 2.7 (-6) &    0.106  \\
01049 & 004 &    1 & 280.831 & -32.783 &  24.66 & 1.3 (+21) & 7.0 (+20) & 8.4 (-6) &    0.314  \\
01050 & 008 &    1 &  89.262 &  30.576 &  19.73 & 3.2 (+20) & 5.3 (+20) & 3.3 (-6) &    0.067  \\
01050 & 010 &    1 &  89.262 &  30.576 &  19.73 & 3.2 (+20) & 5.3 (+20) & 3.3 (-6) &    0.067  \\
01204 & 018 &    1 & 326.055 &  20.828 &  21.92 & 5.4 (+20) & 6.6 (+20) & 4.2 (-6) &    0.158  \\
01204 & 020 &    1 &  22.219 &  19.352 &  19.73 & 1.0 (+21) & 1.8 (+21) & 2.2 (-5) &    0.488  \\
01204 & 021 &    1 & 106.999 & -24.266 &  21.32 & 9.0 (+20) & 4.5 (+20) & 5.5 (-6) &    0.364  \\
01253 & 013 &    1 & 194.390 & -28.832 &  18.63 & 7.2 (+20) & 5.8 (+20) & 7.0 (-6) &    0.151  \\
01261 & 002 &    1 & 271.739 & -32.774 &  18.86 & 3.9 (+20) & 6.2 (+20) & 3.9 (-6) &    0.092  \\
01265 & 002 &    1 &  20.751 &  27.278 &  22.65 & 5.5 (+20) & 4.3 (+20) & 5.2 (-6) &    0.163  \\
01269 & 003 &    1 & 309.626 &  19.409 &  23.89 & 2.7 (+20) & 7.1 (+20) & 8.6 (-6) &    0.245  \\
01282 & 019 &    1 & 278.683 &  22.963 &  19.69 & 4.6 (+20) & 4.8 (+20) & 5.7 (-6) &    0.140  \\
...   & ... & ...  & ...     & ...     & ...    &  ...      & ...       & ...      &       ... \\
\hline
\end{tabular}
\tablefoot{
  Atomic hydrogen column densities, $N(\mathrm{H})$, are derived from the HI4PI survey at an angular resolution of 16.2' \citep{HI4PI2016}. The total hydrogen column densities, $N_{\mathrm{H}}$, are inferred from \textit{Planck} data \citep{Planck2014}, following equation \ref{Eq:Ntot}. The PAH specific intensity at 12~$\mu$m, $B_\nu(12\,\mu{\rm m})$, the average dust temperature, $T_{\rm dust}$, and the dust opacity at 353~GHz, $\tau_{353}$, are obtained from WISE \citep{Meisner2014}, and \textit{Planck} all-sky maps \citep{Planck2014}, both smoothed to 5' resolution. Numbers in parentheses are powers of ten. The full version of this table is available online at the CDS.\\
}
\end{table*}

\subsection{All-sky surveys}

The intensities of the pure rotational lines of \HH\ measured along the lines of sight listed in Table~\ref{tab:all-detect} arise from Galactic material, with systemic velocities and associated uncertainties consistent with the range expected from Galactic rotation. To interpret the origin and excitation properties of this emission, it is therefore essential to estimate the associated gas column densities and dust emission along the same lines of sight. These ancillary quantities provide important contextual information.

To this end, we compiled estimates of the atomic hydrogen column density, $N(\mathrm{H})$, the total hydrogen column density, $N_{\mathrm{H}}$, the PAH specific intensity at 12~$\mu$m, $B_\nu(12\,\mu{\rm m})$, and the average dust temperature, $T_{\rm dust}$, using publicly available all-sky surveys from  HI4PI, \textit{Planck}, and WISE \citep{HI4PI2016,Planck2014,Meisner2014}. Previous combined analyses of HI, dust, and CO emission across the Galaxy have shown that $N_{\mathrm{H}}$ and the dust opacity at 353~GHz, $\tau_{353}$, can be approximated by two distinct linear relations below and above  $N_{\mathrm{H}} \simeq 6 \times 10^{20}$~cm$^{-2}$ \citep{Planck2014}. Following these studies, the total hydrogen column density is computed as
\begin{equation}
\label{Eq:Ntot}
N_{\rm H} =
    \begin{cases}
      \dfrac{\tau_{353}}{6.3 \times \ 10^{-27}}\ {\rm cm}^{-2} & {\rm if}\ \dfrac{\tau_{353}}{8.26 \times \ 10^{-27}} < 6 \times \ 10^{20} \text{ cm}^{-2}\\[0.3cm]
      \dfrac{\tau_{353}}{1.2 \times \ 10^{-26}}\ {\rm cm}^{-2} & {\rm otherwise}.
    \end{cases}
\end{equation}
The resulting values of $N(\mathrm{H})$, $N_{\mathrm{H}}$, $\tau_{353}$, $B_\nu(12\,\mu{\rm m})$, and $T_{\rm dust}$ for each line of sight are listed in Table~\ref{tab:ancillary}.

A key limitation of this approach lies in the spatial resolution mismatch between the ancillary  all-sky surveys and the JWST/MRS observations. The angular resolution of the HI4PI data product used in this paper is 16.2', while that of the dust and infrared maps from \textit{Planck} and WISE is $\sim 5$'. In contrast, the MRS instrument aboard the JWST provides effective FoVs between 6" and 12" depending on the IFU channel (see Sect.~\ref{Sect:Datared}). As a result, the derived column densities and infrared specific intensities represent large-scale averages that may not capture small-scale structure or fluctuations within the MRS FoV. These quantities should therefore be interpreted as coarse contextual estimates rather than precise measurements of the conditions at the exact location of the \HH\ emission.

\subsection{Molecular fraction}

Bearing this caveat in mind, the molecular hydrogen column density along each line of sight is estimated by subtracting the atomic hydrogen contribution from the total hydrogen column density:
\begin{equation}
N(\HH) = \frac{N_{\mathrm{H}} - N(\mathrm{H})}{2}.
\end{equation}
This approach relies on the assumption that fully ionized phases do not contribute significantly to dust extinction, so that the column density derived from dust primarily traces neutral atomic and molecular gas. Because $N(\mathrm{H})$ and $N_{\mathrm{H}}$ are often comparable along the lines of sight studied in this paper—particularly at high Galactic latitude—this procedure frequently yields negative values of $N(\HH)$, reflecting the fact that the molecular component lies within the combined uncertainties of the two quantities. This behavior has been reported in previous studies \citep[e.g.,][]{Skalidis2024}. Given the uncertainties on both $N(\mathrm{H})$ and, more critically, on the calibration used to derive $N_{\mathrm{H}}$, we estimate that $N(\HH)$ cannot be robustly constrained below about 10\% of the total hydrogen column density. However, because the sensitivity of the MRS limits detections to relatively high molecular hydrogen column densities, the  majority of lines of sight with detected \HH\ emission lie above this threshold.

The derived column densities of atomic and molecular hydrogen, the corresponding averaged molecular fraction
\begin{equation}
\label{Eq:mol_frac}
f_{\HH} = \frac{2N(\HH)}{N_{\mathrm{H}}},
\end{equation}
and the PAH and dust specific intensities, span more than four orders of magnitude across the full observational sample. With the exception of a few flagged lines of sight (see Appendix~\ref{append:flag} and Table~\ref{tab:ancillary}), both $N_{\mathrm{H}}$ and $N(\HH)$ remain below $10^{23}$~cm$^{-2}$, thereby validating the optically thin approximation and the omission of dust extinction adopted in Eq.~\ref{Eq:cdint} to convert intensities into upper-level column densities.

\subsection{Selection of diffuse lines of sight}
\label{append:flag}

Not all MRS background observations retrieved from the \href{https://mast.stsci.edu/portal/Mashup/Clients/Mast/Portal.html}{MAST} archive probe purely diffuse interstellar gas. A subset of lines of sight shows enhanced \HH\ rotational emission or elevated dust temperatures, indicative of contributions from irradiated environments such as star-forming complexes or from denser, possibly self-gravitating structures. To isolate sightlines representative of the diffuse ISM, additional selection criteria are required.

\begin{figure*}[!]
    \centering
    \resizebox{\hsize}{!}{\includegraphics[trim = 0.4cm 0.3cm 0.4cm 0.4cm]
    {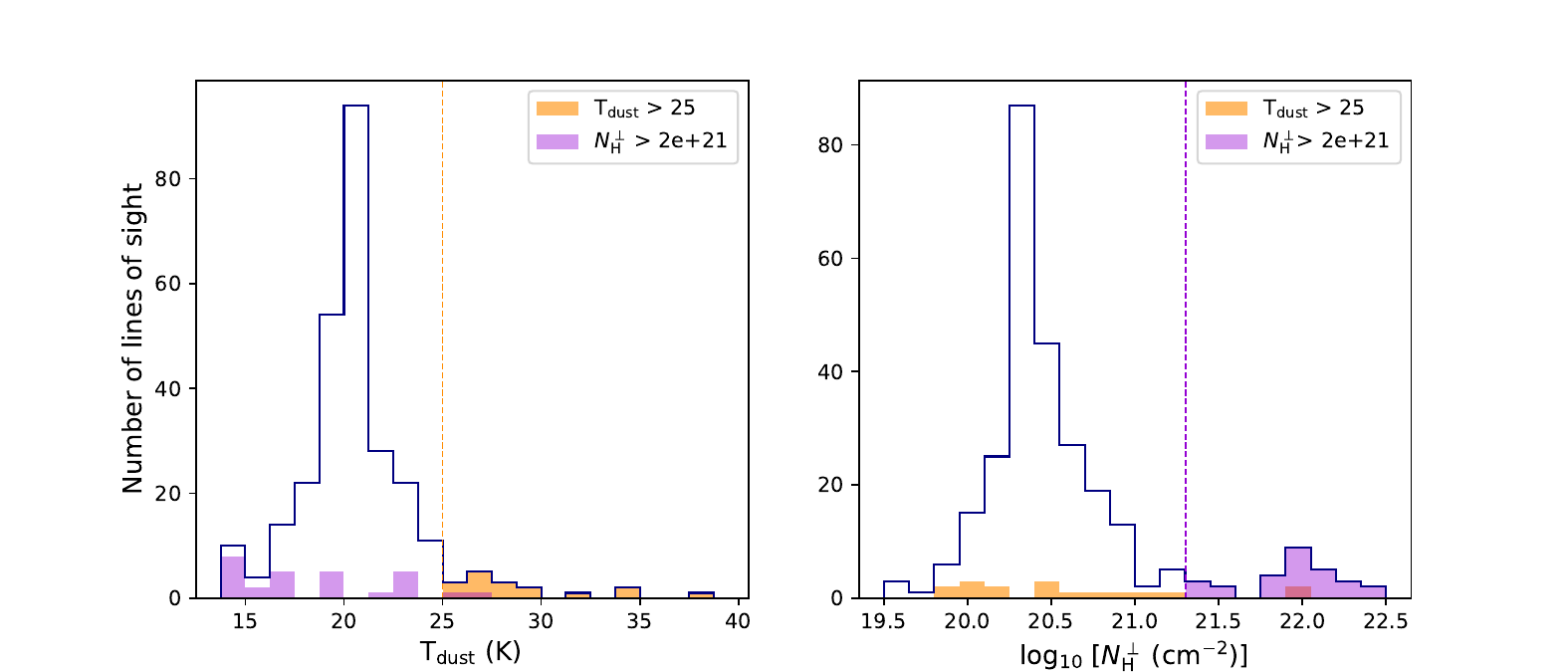}}
    \caption{Integrated properties of the observational sample.
    {\it Left panel:} Distribution of the line-of-sight--averaged dust temperature, T$_{\rm dust}$. {\it Right panel:} Distribution of the logarithm of the total hydrogen column density projected perpendicular to the Galactic plane, $N_{\rm H}^{\perp}$. Lines of sight classified as non-diffuse ($N_{\rm H}^{\perp} > 2 \times 10^{21}$~cm$^{-2}$ or $T_{\rm dust} > 25$~K) are highlighted in color.}
    \label{fig:hists}
\end{figure*}

Figure~\ref{fig:hists} shows the distributions of the logarithm of the total hydrogen column density projected perpendicular to the Galactic plane,
\begin{equation}
N_{\rm H}^{\perp} = N_{\rm H} \sin |b|,
\end{equation}
and of the line-of-sight--averaged dust temperature, $T_{\rm dust}$. The bulk of the sample is distributed in approximately symmetric cores centered around $N_{\rm H}^{\perp} \simeq 3 \times 10^{20}$~cm$^{-2}$ and $T_{\rm dust} \simeq 20$~K, characteristic of diffuse Galactic gas \citep{Miville-Deschenes2017,Planck2014}. A smaller fraction of the lines of sight extends toward higher values of both quantities, forming asymmetric wings indicative of denser or more strongly irradiated environments. We therefore classify as non-diffuse all lines of sight lying more than three standard deviations from the core of either distribution, corresponding to $N_{\rm H}^{\perp} > 2 \times 10^{21}$~cm$^{-2}$ or $T_{\rm dust} > 25$~K.

After excluding these flagged data, the \HH\ column densities associated with detected rotational lines are typically above $10^{19}$~cm$^{-2}$, corresponding to molecular fractions between 0.1 and 1. This indicates that most retained lines of sight have undergone the H--\HH\ transition. Compared to absorption studies in the solar neighborhood, our sample occupies the higher end of the column density and molecular fraction distributions, more representative of sightlines in or near the Galactic plane \citep{Shull2021a}. This reflects the sensitivity threshold of the MRS instrument, which favors the detection of diffuse regions with relatively high molecular content.

\end{appendix}

\end{document}